%% file: main.tex
\documentclass[sigconf,authorversion,nonacm]{acmart}
\usepackage[nameinlink]{cleveref}
\usepackage{graphicx}
\usepackage{xspace}
\usepackage{multirow}
\usepackage{colortbl}
\usepackage{framed}
\usepackage{listings}
\usepackage{xcolor}
\usepackage{adjustbox}

\usepackage[switch]{lineno}
 
\usepackage{ifthen}
\usepackage{xcolor}

\newboolean{showcomments}
\setboolean{showcomments}{true} 
\ifthenelse{\boolean{showcomments}}{
  \newcommand{\nbc}[3]{
    {\colorbox{#3}{\bfseries\sffamily\scriptsize\textcolor{white}{#1}}}%
    {\textcolor{#3}{\sf\small$\blacktriangleright$\textit{#2}$\blacktriangleleft$}}}
  \newcommand{\todo}[1]{\nbc{TODO}{#1}{blue}\xspace}
}{
  \newcommand{\nbc}[3]{}
  \newcommand{\todo}[1]{}
}

\newcommand{\FOUNDRY}{\textsc{Foundry}\xspace}
\newcommand{\prodscalpel}{\textsc{prodScalpel}\xspace}

\begin{document}

\title{Software Product Line Engineering via Software Transplantation }


\author{Leandro O. Souza}
\email{leandro.souza@ifba.edu.br}
\affiliation{%
  \institution{Federal Institute of Bahia}
  \city{Irecê}
  \state{Bahia}
  \country{Brazil}
}

\author{Earl T. Barr}
\affiliation{
    \institution{University College London}
    \city{London}
    \country{UK}
    }

\author{Justyna Petke}
\email{j.petke@ucl.ac.uk}
\orcid{0000-0002-7833-6044}
\affiliation{
    \institution{University College London}
    \city{London}
    \country{UK}
    }

\author{Eduardo S. Almeida}
\affiliation{
    \institution{Federal University of Bahia}
    \city{Salvador}
    \state{Bahia}
    \country{Brazil}
    }

\author{Paulo Anselmo M. S. Neto}
\affiliation{
    \institution{Federal Rural University of Pernambuco}
    \city{Recife}
    \state{Pernambuco}
    \country{Brazil}
    }



\begin{abstract}
For companies producing related products, a Software Product Line (SPL) is a software reuse method that improves time-to-market and software quality, achieving substantial cost reductions. 
These benefits do not come for free.
It often takes years to re-architect and re-engineer a codebase to support SPL and, once adopted, it must be maintained. Current SPL practice relies on a collection of tools, tailored for different re-engineering phases, whose output developers must coordinate and integrate. 
We present \FOUNDRY, a general automated approach for leveraging software transplantation to speed conversion to and maintenance of SPL. 
\FOUNDRY facilitates feature extraction and migration. 
It can efficiently, repeatedly, transplant a sequence of features, implemented in multiple files. 
We used \FOUNDRY to create two valid product lines that integrate features from three real-world systems in an automated way. 
Moreover, we conducted an experiment comparing \FOUNDRY's feature migration with manual effort. We show that \FOUNDRY automatically migrated features across codebases 4.8 times faster, on average, than the average time a group of SPL experts took to accomplish the task.
\end{abstract}





\maketitle
\input{sections/sec01-introduction}

\label{intro}

\input{sections/sec02-motivating_example.tex}

\label{sec:motivating_example}

\input{sections/sec03-approach.tex}

\label{sec:approach}

\input{sections/sec04-implementation.tex}
\label{sec:implementation}

\input{sections/sec06-case_studies.tex}

\label{sec:case_studies}

\input{sections/sec07-experiment.tex}
\label{sec:experiment}

\input{sections/sec05-automated_spl_reengineering.tex}
\label{sec:automated_spl_reengineering}

\input{sections/sec08-related_work.tex}

\label{sec:related_work}

\input{sections/sec09-conclusion.tex}

\label{sec:conclusion} 







\bibliographystyle{ACM-Reference-Format}
\bibliography{main}

\appendix

\end{document}

%% file: sections/sec01-introduction.tex
\section{Introduction}
\label{sec:introduction}

Software Product Line (SPL) is a systematic methodology for producing related software products from shared development assets~\cite{Pohl2005, Linden2007, Fischer2015}.
SPL organises features as core assets that are shared across all products within the product line. Then, it defines a \emph{dependency relation} over those features and a \emph{variability mechanism} that creates products from subsets of the features, subject to the dependency relation.
By centralising features into a product base, SPL prevents cross-product feature drift, allowing all products to benefit from a feature's improvement, if they use it.
For companies producing related products, adopting SPL improves productivity and quality, speeds time to market, and reduces cost, because it facilitates the reuse of development artifacts, such as code and design~\cite{Linden2007,Bastos2015}.

Despite its benefits, adopting SPL requires considerable upfront investment before its benefits can be realised. The cost of migrating existing products to SPL is lower than adopting SPL from scratch, making \emph{extractive}~\cite{Krueger2001} adoption more common, especially in companies with many software system variants in production~\cite{Berger2013}. Two factors drive this preference: 1) it is often hard to know upfront that SPL will be needed because related products often emerge from a small set of initial products, and 2)  starting from scratch discards considerable knowledge and investment in existing codebases, when they exist~\cite{Breivold2008,Northrop2012}.

To re-engineer existing products, companies must solve four problems:  They must analyse their products to 1) identify and 2) extract the features these products share, and 3) learn their inter-dependencies. Finally, they must 4) define a variability mechanism for combining these features, subject to their inter-dependency constraints~\cite{Assuncao2017}. 
Currently, re-engineering to adopt SPL remains largely manual~\cite{Assuncao2017} and costly~\cite{Bockle2004}. Indeed, because of its cost, software companies delay, or even refrain from, adopting SPL~\cite{Fischer2015}.  Automating these tasks remains an open challenge~\cite{Assuncao2017}.


In 2013, Harman et al.~\cite{Harman2013} introduced software transplantation (ST) as a new research direction and laid out its implications for SPL re-engineering. Harman et al. defined software transplantation as ``\emph{the adaptation of one system's behaviour or structure to incorporate a subset of the behaviour or structure of another}''~\cite{Harman2013}. In terms of automated software transplantation, Petke et al.~\cite{Petke2014,Petke2018} were the pioneers in transplanting code snippets from different versions of a system to enhance its performance using genetic improvement~\cite{Petke18}. A year later,  Barr et al.~\cite{Barr2015} introduced a theory, algorithm, and tool that could automatically transplant  a feature from one program to another successfully. Another tool, CodeCarbonCopy (CCC), was proposed by Stelios Sidiroglou-Douskos et al.~\cite{Sidiroglou2017} , which automatically transfers code from a donor to a host codebase by utilizing static analysis to identify and eliminate irrelevant functionalities that are not pertinent to the host system.

Inspired by this line of work, we introduce \FOUNDRY (\Cref{sec:approach}), the first software transplantation approach for SPL re-engineering. \FOUNDRY is independent of the programming language, and supports SPL's \emph{domain engineering} and \emph{application engineering}~\cite{Clements2001} processes at the code level.
It tackles each SPL re-engineering task, easing some and automating others. \FOUNDRY does not eliminate the manual labour of feature identification, but reduces it to the task of annotating the entry points (i.e., the interface) of a feature, or its ``organ'' using transplantation nomenclature. \FOUNDRY amortises this manual step across a sequence of transplantations. 
\FOUNDRY  automates feature extraction; to do so, it uses slicing to overapproximate feature dependencies.  It leverages transplantation to automate the variability mechanism and, simultaneously, tackle slice-imprecision.
Key to \FOUNDRY is mapping software transplantation's ``over-organs'', conservative program slices~\cite{Barr2015}, to product line assets, or features. 
Its use of slicing means that \FOUNDRY does not need specially prepared donors; The donor programs can even be unaware that they are participating in an SPL. A product line via ST is composed of multiple over-organs and a ``product base'', a host that contains all features are shared across all products within the product line, so constructing a product entails transplanting a set of organs into a product base. 

Because over-organs are conservative, self-contained slices, two organs may share features. For example, in an editor, two different features, like a spell checker or a plugin manager, might share a memory-resident database feature. \FOUNDRY uses clone-aware genetic improvement~\cite{Petke18} to specialise an over-organ to its implantation point and to detect and remove cross organ redundancies (\Cref{sec:approach}). We realise \FOUNDRY in \prodscalpel, a tool that transplants multiple organs (i.e., a set of interesting features) from donor systems into an emergent product line for codebases written in C.  
\prodscalpel also supports the use of existing variability mechanisms~\cite{Gacek2001} based on \emph{feature toggle}~\cite{Rahman2016} or \emph{preprocessor directives}~\cite{Kastner2008B}. It can surround implanted organs with feature flags, which permit enabling and disabling features, to facilitate its integration into an existing SPL codebase that uses them. 

To evaluate \prodscalpel, we conducted two case studies (\Cref{sec:case_studies}) and a controlled experiment (\Cref{sec:experiment}). We first generate products by transplanting features from three real-world systems ---  Kilo\footnote{https://github.com/antirez/kilo}, VI\footnote{http://ex-vi.sourceforge.net/} and CFLOW\footnote{https://www.gnu.org/software/cflow/} --- into two product bases generated from VI and VIM\footnote{https://www.vim.org/}, used as hosts for the target transplantations. Next, we conducted an experiment in which we asked twenty SPL experts to move a feature into a product line. 
We gave them the same inputs as those \prodscalpel requires. In all cases, \prodscalpel outperformed our experiment's participants in the time taken to transplant the feature.
On average, \prodscalpel took 18\% of the time to transplant features from single systems to a product line than the participants who completed the task within the timeout.

Our results (\Cref{sec:result_discussion} and  \Cref{sec:evaluation_results}) show that software transplantation speeds SPL re-engineering, by combining features extracted from existing, possibly unrelated, systems.

The main contributions of this paper are: 
\begin{enumerate}
    \item \FOUNDRY, a novel SPL re-engineering approach that leverages software transplantation to extract and reuse features from existing codebases to construct a product line, even when those features and their codebases were not built for, or even aware of, the product line.

    \item \FOUNDRY's realisation for C in \prodscalpel, a  tool that transplants multi-file organs and uses clone detection to prevent implanting redundant features;
    
   \item A rigorous evaluation of \prodscalpel that demonstrates \FOUNDRY's promise. We use \prodscalpel to generate two product lines and two new products, composed of features transplanted from three different real-world codebases. We also show that \prodscalpel migrates features on average 4.8 faster than SPL experts performing the same task.

\end{enumerate}

 All the source code and data needed to reproduce this work are available at the project webpage~\cite{ProjectWebpage}.

%% file: sections/sec02-motivating_example.tex
\section{Motivating Example} \label{sec:motivating_example}

 \begin{figure*}[t]
	\centering \includegraphics[width=\textwidth]{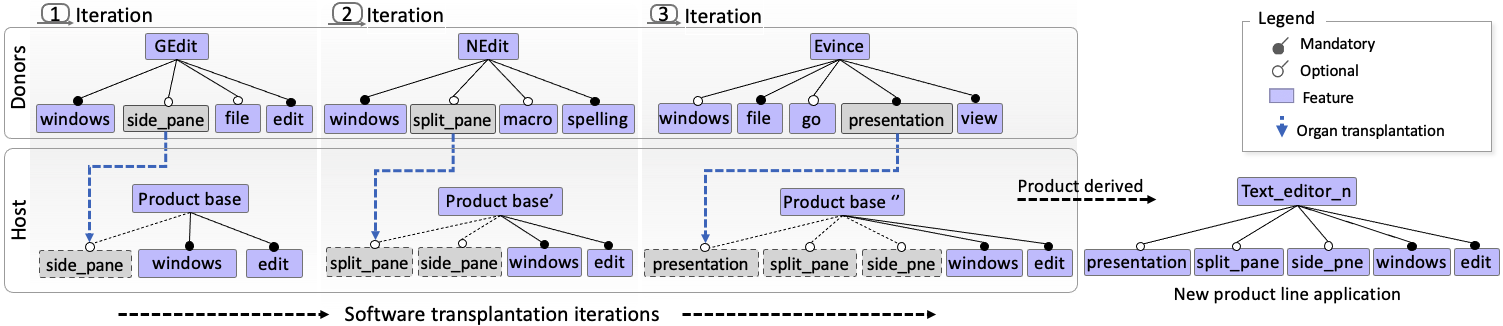}
	\caption{Product derivation process using the \FOUNDRY approach. \textit{\prodscalpel transplants three features, in sequence, into the GEdit's product base to derive a new text editor after three iterations of organ transplantation.} }
	\label{fig:incremental_pd}
\end{figure*} 

The open-source GNOME project\footnote{https://wiki.gnome.org/Projects} encompasses a large portfolio of individual programs evolve as independently as possible from the rest. These programs share features, but because they are separately developed, their constituent features cannot be easily reused across its portfolio to provide mass customization, at least without much manual effort. The combination of mass customization and a common platform, principles of SPLE~\cite{Pohl2005}, would allows to GNOME team reuses a common base of technology and, at the same time, to bring out products tailored individual customers. Without a common platform and a software development process base on mass customisation, it may be more difficult to the GNOME project provides customized products and effectively manage the commonality and variability of its features.
The GNOME  project is a natural candidate for SPL, but the significant re-engineering investment of time and resources have prevented it from adopting SPL. \FOUNDRY is transformative in this case because it can be used to reduce this cost. By using \prodscalpel for automated support, the GNOME team can iteratively and incrementally reengineer the codebases of GNOME's application portfolio for SPL. 

Suppose project collaborators want to build a product line in the domain of text editors. This product line would allow GNOME to produce text editors that augment its current text editor, \emph{GEdit}\footnote{https://wiki.gnome.org/Apps/Gedit}, with additional features. Since they have decided to augment Gedit, GNOME team would select it as the product base, the shared substrate of a product line that, for \FOUNDRY, serves the host for transplanted features. Assume that the GNOME team targets the following three features (1) $\texttt{side-panel}$, (2) $\texttt{split pane}$, and (3) $\texttt{presentation}$. They then identify two donors from which to transplant these features:  \emph{NEdit}\footnote{https://sourceforge.net/projects/nedit/}, a multi-purpose text editor that is not part of the GNOME portfolio, and \emph{Evince}\footnote{https://wiki.gnome.org/Apps/Evince}, a document viewer for multiple document formats that is part of GNOME, not not an editor. 
 
Once defined all possible donors and a host, the GNOME engineers can start the process. In the donors, they need to demarcate all feature entry points into transplant; a single annotation is sufficient for \prodscalpel to extract a feature. To prepare the host, GNOME engineers use \prodscalpel to extract a product base from an \textit{GEdit} by removing all features not shared across all products within the built product line. 

To demarcate the inserting point in the product base, the engineers must annotate it to indicate the implantation point for each target feature, or “organ” using transplantation nomenclature. GNOME engineers then run \prodscalpel on these inputs, once per feature, with 


 
Then, the engineers must annotate the host to indicate the implantation point for each target feature, or “organ” using transplantation nomenclature. GNOME engineers then run \prodscalpel on these inputs, once per feature, with 
\begin{quote}
    $\texttt{./prodScalpel \textemdash \textemdash seeds\_file}$: \textit{ The file which contains the seeds for Genetic Programming(GP) algorithm.} 

    $\texttt{\textemdash \textemdash donor\_folder}$: \textit{The path to the donor source code.} 
    
    $\texttt{\textemdash \textemdash host\_target}$: \textit{The file in host that contains the insertion point of the transplant.}
    
    $\texttt{\textemdash \textemdash donor\_target}$: \textit{The file in the donor that contains the core function.}
    
    $\texttt{\textemdash \textemdash workspace}$: \textit{The path to the workspace of the transplant.}
    
    $\texttt{\textemdash \textemdash core\_function\_target}$: \textit{The file which contains all feature entry points.}
    
     $\texttt{\textemdash \textemdash  host\_project}$: \textit{The path to the product base source code.}
     
   \end{quote} 
%
%

\prodscalpel automatically extracts all of the specified feature's source code and its dependencies, or “over-organ” using transplantation nomenclature. More operational parameters are available at the project: webpage~\cite{ProjectWebpage}.

\Cref{fig:incremental_pd} illustrates all transplantation iterations performed to generate a new product.  It shows a new text editor derived from the transplant of features from different donors and using GEdit as a product base. Using feature models~\cite{Kang1990} to represent each donor system and the product base evolution, \prodscalpel first transplants the \emph{side-panel} feature, extracted from GEdit itself. This transplantation demonstrates that \prodscalpel can transplant features into a product base that comes from the same codebase.  Next, \prodscalpel transplants the \emph{split\_pane} feature from NEdit. It shows how \prodscalpel manages to transplant features from distinct codebases, which is not possible without manual effort using the current state-of-art to SPL re-engineering. Finally, \prodscalpel transplants the \emph{presentation} feature from GNOME's Evince renderer. 

\FOUNDRY facilitates transplanting features from any program into a product line, opening the door to large scale feature reuse. Open-source projects, like GNOME, are an especially promising source of code for \FOUNDRY, so long as the donors and target hosts share compatible licenses.

%% file: sections/sec03-approach.tex
\section{Foundry} 
\label{sec:approach}

Based on software transplantation idea, \FOUNDRY treats \emph{product base} and \emph{over-organs} (representing features) as product line assets. 
A product base is a host that contains all features that will be shared among the products. 
An over-organ, in turn, is a completely functional and reusable portion of code extracted from a donor system that conservatively over-approximates the target organ~\cite{Barr2015}. An over-organ can be specialized to became an organ that preserves the original behavior of the feature in a different host codebase~\cite{Barr2015}.

\begin{figure*}[t]
	\centering \includegraphics[width=0.8\textwidth]{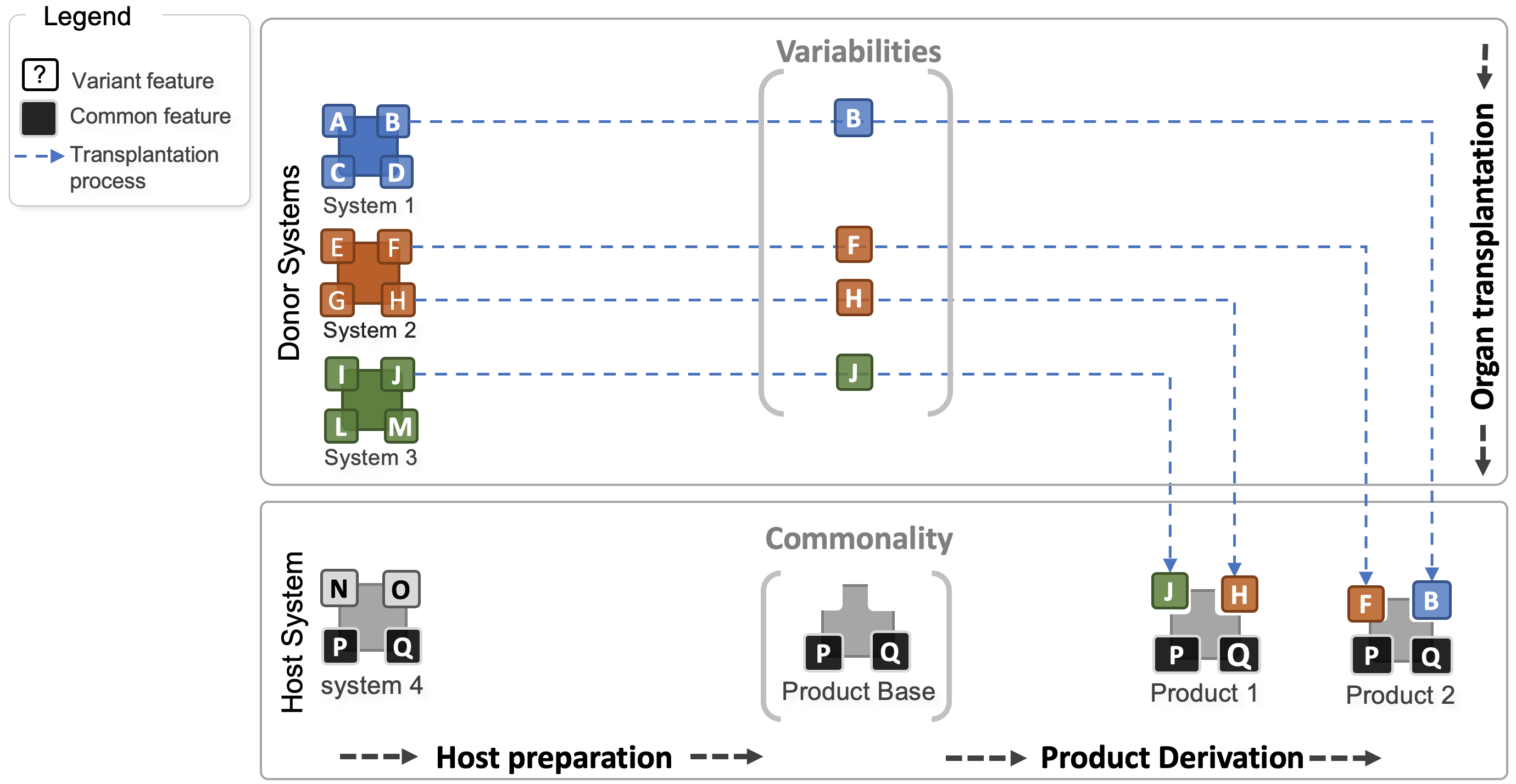}
	\centering 
	\caption{An overview of how new products are derived from a product line based on ST.}
	\label{fig:product_for_transplantation}
\end{figure*}

Conceptually, in \FOUNDRY, while the product base provides commonalities (i.e., common features) to the target product line, the variability (i.e., variant features) are provided by the transplantation process, as illustrated in \Cref{fig:product_for_transplantation}. This idea opens new ways for SPLE area by automated construction of different products by transplanting multiple organs into a product base. 

It is important to note that \FOUNDRY offers two ways of creating products by transplanting organs from a pre-established/created \emph{transplantation platform}, a repository of transplantation assets, or by directly extracting and transplanting an organ from a donor into a product base, even if the over-organ is not present in the transplantation platform. The two ways can also be combined to create specialized products.

In the rest of this section, we overview \FOUNDRY's workflow, describing how it applies software transplantation idea to re-engineering of product lines from existing systems. \FOUNDRY is independent of the programming language, and supports SPL's \emph{domain engineering} and \emph{application engineering}~\cite{Clements2001} processes at the code level.

\subsection{Domain Engineering}

In SPL's lifecycle the domain engineering corresponds to the process of establishing a reusable platform of core assets~\cite{Clements2001}. The process defines what will be shared among the products derived from it, i.e., commonalities. It also specifies the possible variations expressed as artefacts, that will enable the customization of product line applications, i.e., products.

\begin{figure*}[t]
	\centering  \includegraphics[width=0.8\textwidth]{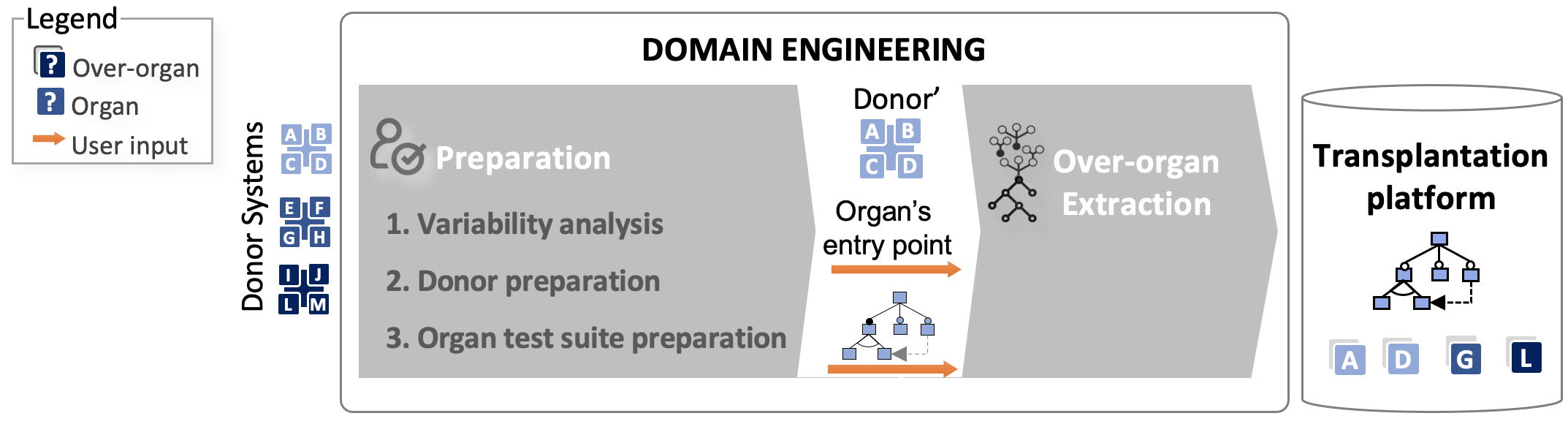}
	\centering \caption{Domain engineering process supported by FOUNDRY. \textit{Four over-organs (A, D, G, L) are extracted from three donor systems and kept in the transplantation platform, with product base consisting of 2 features shared across all products (P and Q).} }
	\label{fig:foundry_dom}
\end{figure*}

In \FOUNDRY, domain engineering corresponds to the process of establishing a product line composed of a product base and a set of reusable over-organs extracted, both stored in the transplantation platform. \Cref{fig:foundry_dom} illustrates the stages of the domain engineering process. 

As in medicine, \FOUNDRY has a \emph{preoperative} stage where donors and the host are prepared for the transplantation process and a \emph{postoperative} stage where we evaluate if the transplantation was successful (\Cref{sec:application_eng} for more details on the postoperative stage). The preoperative stage defines pre-transplantation tasks, responsible for the variability analysis process, the organ's test suite, donor and host preparation.

\textbf{Variability Analysis.} The preoperative stage starts with a variability analysis process to discover features in existing products with the potential to create the target product line. This process aims to create a variability model to express the valid combinations of features among the donor systems from which they will be extracted.  

A variability model can be represented by using a feature model~\cite{Kang1990}. We augmented the traditional feature model representation to incorporate software transplantation inputs required to manipulate and maintain the over-organs in the product line. Each over-organ representation in the feature model is annotated with its corresponding entry point in the donor codebase. 
An organ's entry point is a function in the donor system that belongs to the organ, defines an execution environment expected for its initialization, and provides access to the organ's test suite~\cite{Barr2015}. To determine the organ’s entry point, the SPL engineer needs to provide the name of the function implemented in the donor codebase.

\textbf{Donor Preparation.} This task in the preoperative process consists of cleaning up of donor codebases.  
These can contain some code that will never be used in the target products.
For example, codebases written in C, in general, have code fragments guarded by \#ifdef C-preprocessor directives~\cite{Tartler2011})  commonly used to control code extensions related to features. Although useful for the donor program, such code, if transplanted as part of the target organ, will generate dead code ~\cite{Tartler2011} that will never be executed in any transplanted product.  

Previous work~\cite{Barr2015} with focus on transplantation of a single feature did not concern with donor clean-up. 
However, even when transplanting a single organ, dead code, if not removed apriori, can lead to unnessasary bloat and lower efficiency of the over-organ adaptation process (\Cref{sec:application_eng}). 
In \FOUNDRY, the donor clean-up can be performed in a manual or automated way. During the process the source code structure of the program needs to be preserved (indentation, spacing, number formats, etc.), to prevent future bugs. 

\textbf{Organ Test Suite Preparation.} For product derivation, an SPL engineer must supply test suites, called \emph{ice-box tests}~\cite{Barr2015}. They are used to guide genetic programming in the over-organ adaptation process to create an organ that is fully executable when implanted in the product base (see \Cref{sec:organ_reduction}). 
Ice-box tests can be easily implemented as proposed by~\cite{Barr2015} and integrated into the transplantation platform to be used for new transplantations of the target organ. 
These can be quick to develop, using existing test generation tools, or even adapted from donor’s unit tests, when available.

Even though the preparation process may require manual effort for identifying features of interest,  localising and removing  dead code from the donor and preparing all test suites for the organ, it can be amortised across multiple transplantations and reuse of a single over-organ.

\textbf{Host preparation.} The SPL engineer has to select a product base. It is an existing system which already provides a set of solutions (features) so close to the target products that it can be used as a baseline for the assembly of products. For example, a text editing program could provide a baseline for new programs for text translation,  presentation or rendering, since they could have a considerable number of common features between them.

In case it is necessary, the product base can be reduced to its basic form, keeping only mandatory or features relevant to the target product line. The reduction process consists of removal of all code that implements all features which will not be required to create the target products. In a removal scenario, an additional attention is necessary to find and remove all portions of code that implement each unwanted feature without compromising the product base.

A product base, once reduced, can be used multiple times as a base for new products. Although its preparation may require considerable effort for localising and removing all unnecessary code, it can be compensated with the benefits achieved through using the product base as a baseline to build other products belonging to the same or similar domain.

\textbf{Over-organ Extraction.} Once the donor is prepared, it is possible to start the over-organ extraction process. In this stage,  all source code related to the organ to be transplanted, that implements the target feature, must be identified and extracted. 

Conceptually, an over-organ is composed of an organ and its vein, in keeping with the transplantation analogy~\cite{Barr2015}. While the organ implements the functionality we wish to transplant, the vein is a path from the donor entry that builds and initializes an execution environment for the organ~\cite{Barr2015}. Thus, all the code belonging to the organ and its vein must be captured in the extraction process.

In practice, organs can consist of several lines, one or more classes, as long as they fully implement a specific functionality~\cite{Wang2018}. We consider an organ to be a functional implementation of a feature of interest to an emergent product line. 
This makes the code extracting process a challenging task since it also involves identifying all code elements for the organ to be kept functional even out of the donor's environment.
The extraction of a target over-organ can thus involve a considerable amount of code at different levels of granularity, from moving required files and libraries to entire functions and individual statements, both potentially not confined to a single class, file or library~\cite{Wang2018}. For instance, the feature $\texttt{FEAT\_DIFF}$ implemented in VIM has more than 5k LOCs scattered across 33 of its 166 source files. 
Previous work extracted features from a single file~\cite{Barr2015}.

Here we have four challenges that need special attention. 
First challenge concerns the integrity of the extraction when the extracted code appears in several files, issue also identified by Wang et al.~\cite{Wang2018}. 
In this case, the extracted organ also needs to preserve the original multi-file structure. 
Otherwise, it may be even more difficult to keep it functioning outside of  the donor's environment or to propagate eventual changes to the desirded feature (bug fixes, enhancements, etc.).
Second, defects might be introduced due to code redundancy stemming from multiple organ transplantations. 
Multiple organs can use the same code, creating code duplication and possible errors, if not handled correctly.
Third, an organ's vein can contain a large amount of code, especially when the host and the donor have very different structures. 
This requires extensive modification during the adaptation process. 
According to Barr et al.~\cite{Barr2015},  inlined function calls constitute the second largest number of code lines transplanted. 
Fourth challenge is that an extracted over-organ may itself contain multiple smaller features, which its functionality depends on.
For example, a $\texttt{spell\_checker}$ feature might depend on a memory-resident database feature. 
Thus, the extraction process has to implicitly learn feature's dependencies, by including them in its over-organ. 
Such issues, if not solved, make impracticable the generation and maintenance of product lines using the software transplantation approach. 

To solve all additional challenges highlighted above, we have evolved the organ extraction process, introduced by Barr et al.~\cite{Barr2015}, to compute slices in multiple files. 
Thus, even if an over-organ is contained in multiple files, \FOUNDRY can obtain a practical over-organ for transplantation, maintaining its original structure. 

Given an entry point in the donor provided by the user, \FOUNDRY uses conservative slicing to automatically extract a feature into an ``over-organ'', completely automating the extraction task. Based on the method introduced in \cite{Barr2015}, \FOUNDRY slices \emph{forward} and \emph{backward} from the given organ entry point to identify the organ and one of its veins. 

At the end of the extraction process, the  source code of the over-organ is then stored in the transplantation platform, together with other over-organs that compose the product line. 
All over-organs in the platform are available to be reused during the application engineering process. 

\subsection{Application Engineering}
\label{sec:application_eng}

In SPL, \emph{application engineering} corresponds to the phase where features are assembled to create a product. This is the phase where variability is realized so that artefacts customization takes place.

In \FOUNDRY, application engineering corresponds to the phase of developing customized products through the organ transplantation process. 
After the execution of multiple iteractions of organs transplantation a new product is derived as new organs are transplanted. Thus, the flexibility required to customize products is provided by extracted over-organs that are combined with a product base.

\begin{figure*}[t]
	\centering  \includegraphics[width=0.8\textwidth]{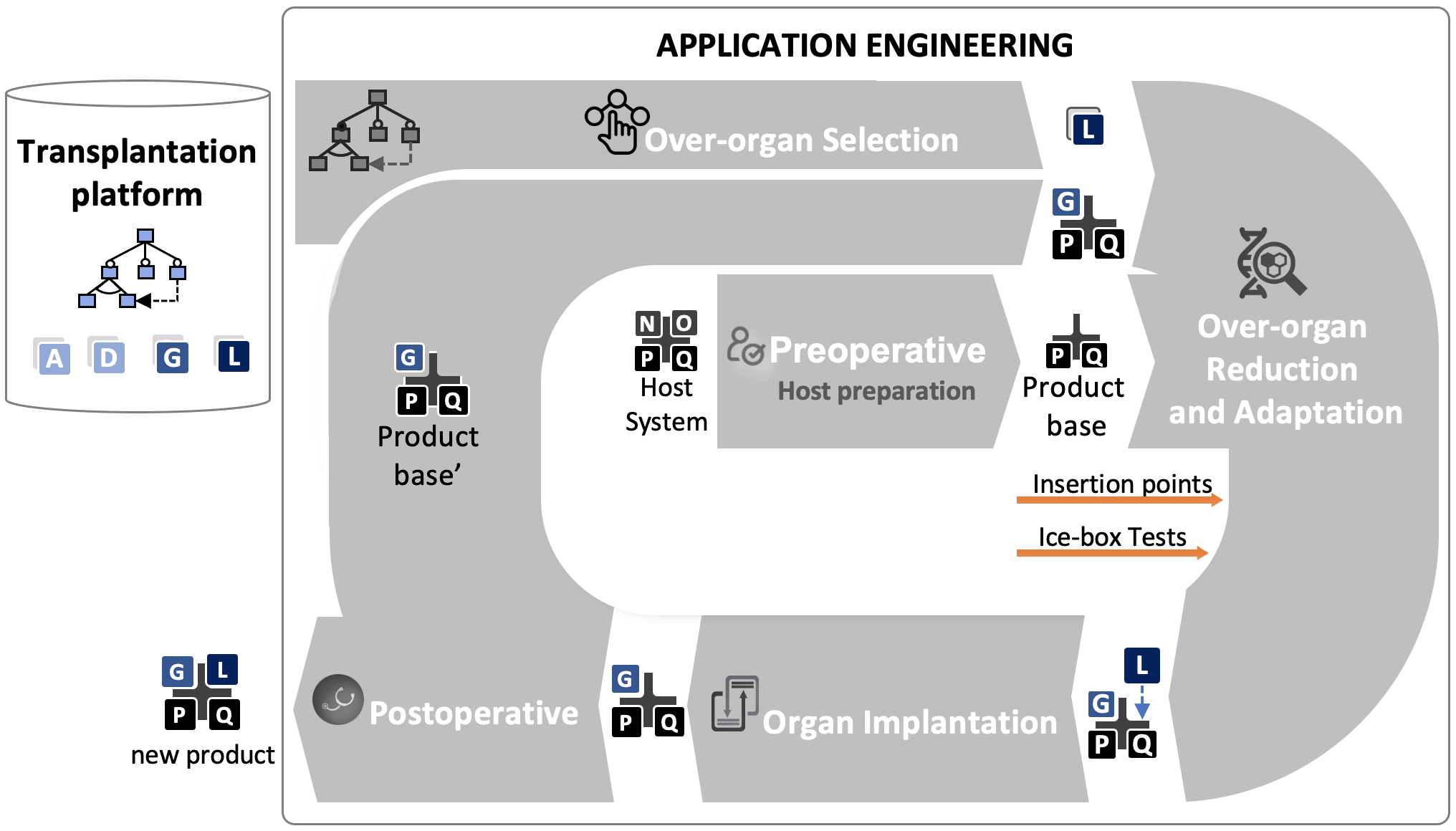}
	\centering \caption{Application engineering process supported by FOUNDRY. \textit{A new product is derived after two software transplantation iterations (organs G and L).}}
	\label{fig:foundry_app}
\end{figure*}

As illustrated in \Cref{fig:foundry_app}, application engineering process is supported by \FOUNDRY by applying four stages of software transplantation: (i) over-organ selection, (ii) over-organ reduction and adaptation, (iii) organ implantation and (iv) postoperative stage. 

\textbf{Over-organ Selection.} At the start of the transplantation process, an SPL engineer selects the target features that will be transplanted into the product base to create the target product. 
The choice is guided by the feature model generated during the variability analysis process, so as, to support the SPL engineer to handle eventual relationship and restrictions among transplanted organs. Once the target feature is chosen in the feature model, it is possible to find the corresponding over-organ in the transplantation platform.

\textbf{Over-organ Reduction and Adaptation.} The over-organ reduction and adaptation processes involve specialising the organ to the host environment, i.e., the product base. An SPL engineer must select the target product base with an annotated implantation point where a call to the organ will be grafted to initialize and execute it. 

Barr et al.~\cite{Barr2015} automated the over-organ reduction and adaptation process using a GP algorithm.
GP reduces an over-organ and specialises it to the host environment.
It thus creates an organ that preserves the original behaviour of the feature at a given insertion point in the host environment. 
However, Barr et al.'s approach does not support the adaptation of an organ that contains multiple files. 
Moreover, it does not support organ maintenance tasks, but rather provides a one-off transplantation approach.
To overcome these issues \FOUNDRY introduces an organ-host wrapper.
This layer is responsible for providing access to the organ from the target host. 
It is automatically constructed on demand, according to a given implantation point in the product base. 

The organ-host wrapper frees developers from the burden of manually writing the code to convert host's data structures into parameters to the organ's entry point whenever a new product is demanded from the product line.

\label{implantation}
\textbf{Organ Implantation.} In this stage of the process, the organ is ready to be implanted into the product base. 
However, to make the application of this technique in generating SPL feasible, we have to consider the transplantation of multiple organs into a single host, a product base, including ones extracted from the same donor. 
As a consequence, the process is no longer concerned with a single but multiple organs and its consequent dependency and interactions.

Feature dependency is a well-known problem in software reuse~\cite{Ribeiro2011}. Dependencies among features are established by means of structural dependencies in the source code shared between elements of different features~\cite{CafeoA2016}. In practice, an organ implementing a feature in a system often shares elements, such as variables and functions with one or more organs that belong to the same donor. For instance, a structure that stores data that are manipulated by more than one function or file; or a function call between the code that belongs to organ A and B. When this happens, the \emph{organ collision} problem occurs since both organ A and B extracted from the same donor must contain the shared function. 

\begin{figure}[t]
	\centering \includegraphics[width=\linewidth]{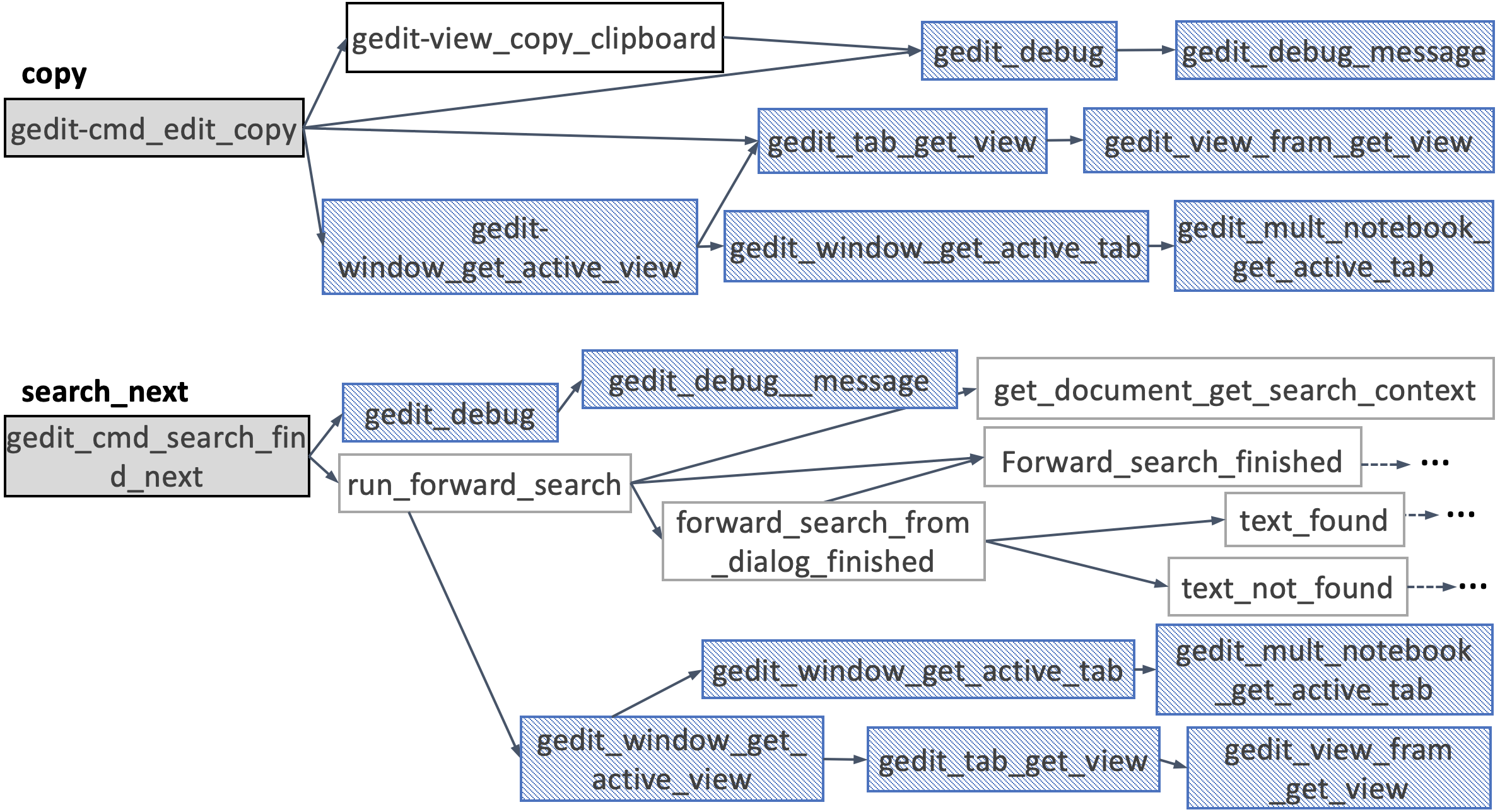}	\caption{Call graph extracted from GEdit text editor. \textit{An example of connection points among call graphs from organs $\texttt{copy}$ and $\texttt{search\_next}$. Highlighted with blue boxes are functions belonging to both organs.}}
	\label{fig:organs_connection_point}
\end{figure}

\Cref{fig:organs_connection_point} shows a real-world example of two call graphs from the same donor, GEdit text editor, sharing several functions. If we consider them as part of two unrelated organs, all common functions (highlighted by blue boxes) will be duplicated during their corresponding implantation processes. As a consequence of organ collision, transplantation process can insert code that is duplicated from multiple organ transplantations. Such a problem, if it is not managed, will lead to unwanted duplicated code, possibly breaking the postoperative product. 

In \FOUNDRY, code elements (functions, directives, constants, declarations of several global variables and their definitions) already belonging to the beneficiary or to more than one organ are characterized as \emph{implicit connection points} since they can represent a connection or dependence points among two or more organs. 

To correctly work, the implantation process need to identify all duplicated code elements and insert them only once into the product base, avoiding code duplication. However, hosts tend to have large input spaces into which code is inserted. In this way, finding the implicit connection points in the host can be difficult. For instance, functions can have the same namespace but not be identical. Thus, it is necessary to check whether a specific code element is already present in the host, considering not only its namespace but its structure and context at a fine level of granularity to make sure that two portions of code are "clones".

Such particular aspect represents a new challenge in the software transplantation field for SPL, handled by \FOUNDRY. 
We solve this challenge by using code clone detection, to avoid duplicated code insertion.

\begin{figure}[t]
\centering \includegraphics[width=8.5cm]{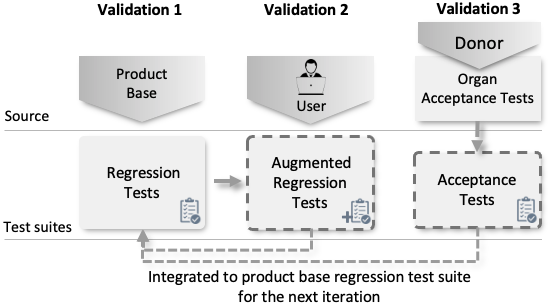}
\caption{Three validation steps; the dashed boxes are test cases added into the host regression test suite after each transplant iteration.}
\label{fig:postoperative_tests}
\end{figure}

\textbf{Postoperative Stage.} As in medicine, \FOUNDRY requires checking the side-effects of the transplantation operation. Based on \cite{Barr2015}, \FOUNDRY's postoperative stage introduces three validation steps, as outlined in \Cref{fig:postoperative_tests}. Extending the validation process proposed in~\cite{Harman2013, Barr2015}, we highlight the three test suites that \FOUNDRY uses to evaluate the quality of a transplant: \emph{Regression}, \emph{Regression++}, and/or \emph{Acceptance} tests.
Once selected, organ has been successfully transplanted, and the postoperative product has passed all postoperative validation steps, we incorporate the Regression++ and Acceptance tests into the host's existing regression test suite for use in the next transplantation.

After the postoperative stage, new iterations of organ transplantation can be performed; thus, in a stepwise and incremental way, a new product is derived as organs are transplanted into a product base.

%% file: sections/sec04-implementation.tex
\section{Implementation} 
\label{sec:implementation}

We now present \prodscalpel, our realisation of \FOUNDRY for C. \prodscalpel extends $\mu$Scalpel~\cite{Barr2015} to transplant multiple organs at once and implant them into a single host codebase. This feature is essential for supporting \FOUNDRY, since many features, especially those existing, SPL-oblivious codebases, span multiple files.

\begin{figure*}[t]
	\centering  \includegraphics[width=\textwidth]{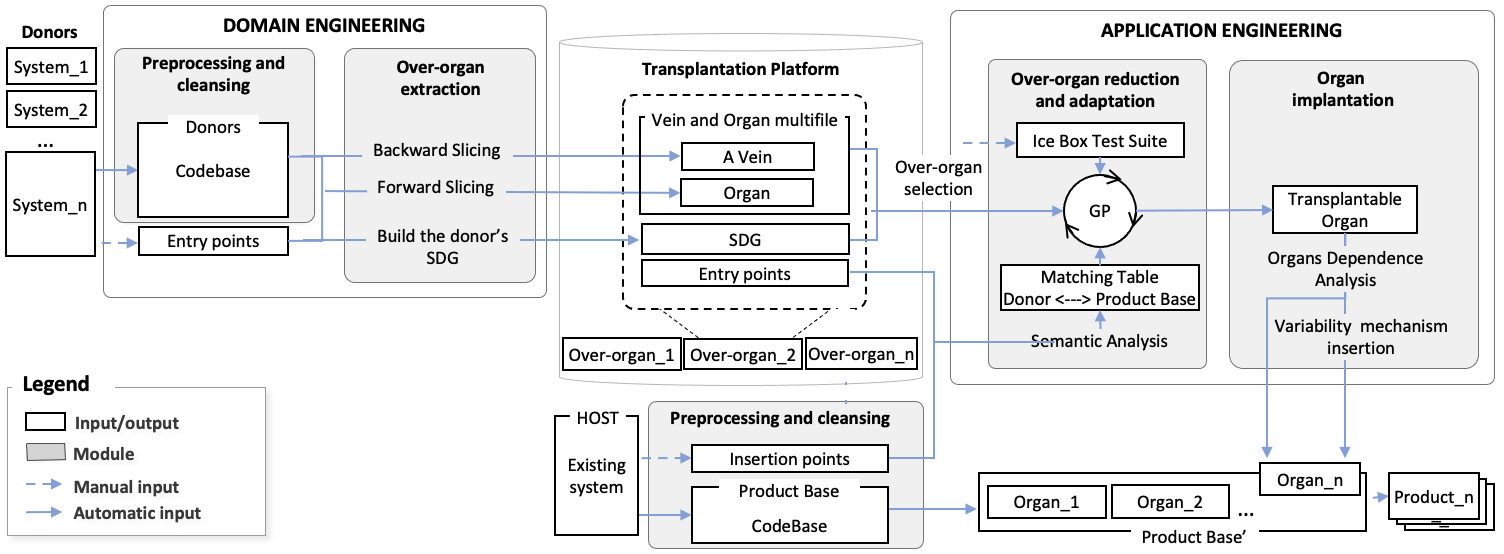}
	\centering \caption{Overall architecture of \prodscalpel. \textit{An SDG is a system dependency graph that the genetic programming phase uses to constrain its search space.}
 }
	\label{fig:prodscalpel}
\end{figure*}

The \prodscalpel's overall architecture is composed by five modules as shows in \Cref{fig:prodscalpel}. The domain engineering process is automated by the \textit{Preprocessing and cleansing} and the \textit{Over-organ extraction} modules. The first one is responsible for cleansing the donor codebases, eliminating any extraneous preprocessor directives that may be present. It also used to host preparing process by removing undesired features delimited by preprocessor directives from the product base. On the other hand, the \textit{Over-organ extraction} module employs the program slicing technique implemented with a System Dependency Graph (SGD) to extract an over-organ from the donor codebase, thereby isolating the relevant code fragments for further adaptation.



The application engineering process is automated by the \textit{Over-organ reduction and adaptation}, and the \textit{Organ implantation} modules. The first one utilizes GP to reduce the size and complexity of the selected over-organ obtained from the transplantation platform. This reduction process is carried out while ensuring that the over-organ is effectively adapted to function within the target product base. GP enables the systematic exploration and optimization of the organ's code structure, achieving an organ specialized to a specific product base. The \textit{Organ implantation}, in turn, employs a clone detector to identify code elements duplication and dependencies during the process of implanting the organ into the product base.

More details on how \prodscalpel automates the \FOUNDRY process is provided in the following sections.

\subsection{Automating Feature Removal}



\begin{figure}[t]
	\centering \includegraphics[width=8.6cm]{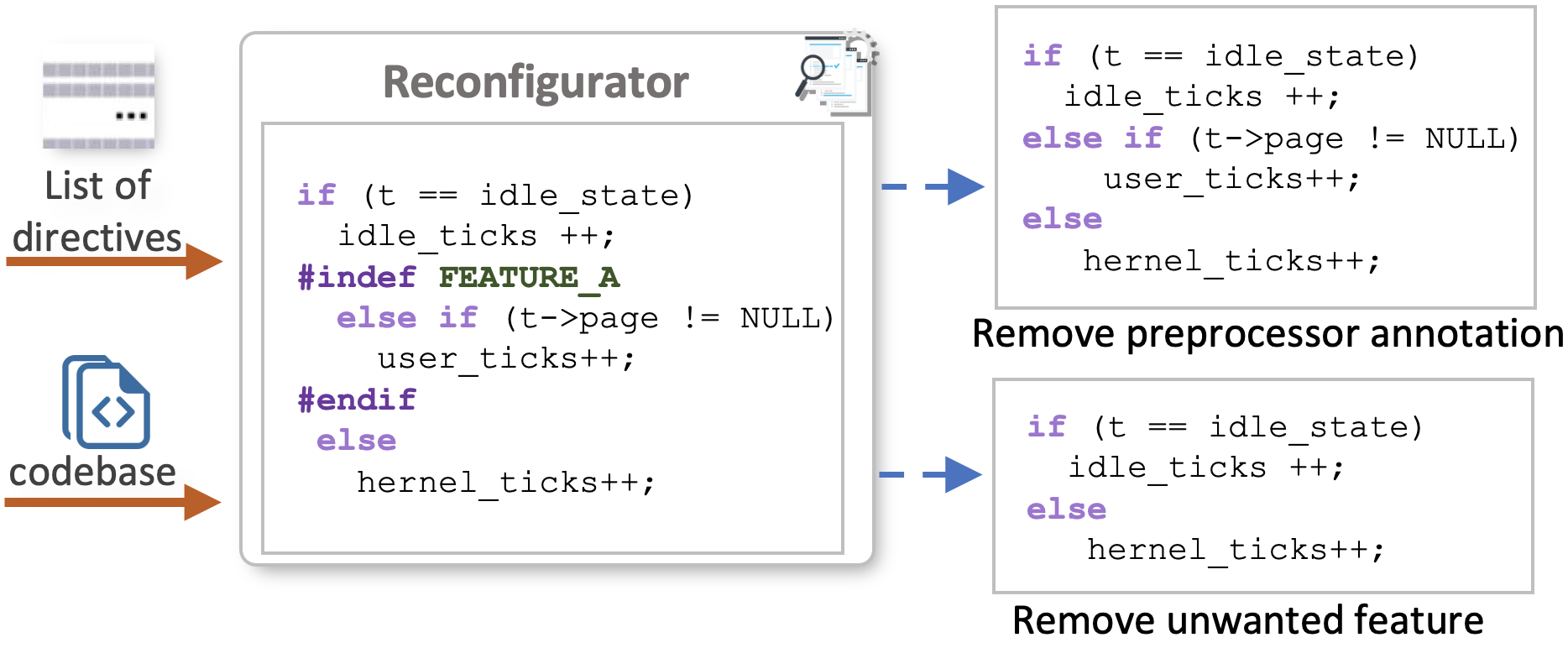}
	\centering 
	\caption{The reconfigurator used to automated removal of preprocessor annotation from donor codebase and unwanted features from product base.}
	\label{fig:reconfigurator}
\end{figure}

First we need to select and prepare our product base. The selected program might contain some unwanted features for the target product, thus we want to remove them before transplantation takes place. To automate removing unwanted features and dead code from the donor and host systems, \prodscalpel  implements a  \emph{Reconfigurator}. To work, \prodscalpel initially requires engineers to provide as input a textual list of preprocessor directives corresponding to each feature and annotation to be removed. From this, the tool searches for pieces of code that implement features limited by preprocessor directives, removing them from the product base. \prodscalpel then maps and searches selectively for pieces of code that implement features limited by such directives. Then, \prodscalpel removes all features source code from the product base while it keeps the source code structure belonging to the product base unchanged and ready to receive the transplanted organs. \Cref{fig:reconfigurator} gives a example of a portion of code after \prodscalpel cleaned up unused directives. 

\subsection{Automating Vein and Over-Organ Extraction}
As optimisations, \prodscalpel extracts the vein from multiple files. \prodscalpel uses backward slicing to identity each function belonging to the vein and saves it in a copy of its souce file in the transplantation platform. Then, it includes all functions calling existing into vein source code in the \emph{array of over-organ statements} used by GP to achieve an organ from the over-organ~\cite{Barr2015}. Thus,  a function/method in the vein shared between two organs is not duplicated when transplanted.

\prodscalpel avoids the problem of vein redundancy when a great part of the vein is common for multiple organs transplanted from the same donors. The vein duplication is discovered by the implantation module during the implantation stage. In the extraction stage, both veins belonging to two organs (A and B) are kept duplicated in the transplantation platform in different directories, since it is important to keep the over-organ functional. However, when the second organ( organ B) is implanted after the transplant of organ A, the code clone detector looks for duplication also in the vein code of organ A. For example, imagine that a vein has a function $\texttt{fx()}$ implemented in file $\texttt{F\.c}$ and belonging to both over-organs A and B. When \prodscalpel tries to transplant organ B after organ A, it checks if the function  $\texttt{fx()}$ already exists in file F.c in the post-operative environment. In this way, if the function  $\texttt{fx()}$ is already in the host post-operative, as part of organ A, \prodscalpel does not insert it into the host again. Instead, the function $\texttt{fx()}$ already transplanted also is used by organ B.

\prodscalpel also identifies occurrences of mutually recursive functions, even an occurrence of an indirect recursion. Technically, \prodscalpel inlines the vein code while puts each function found in a stack of functions. When \prodscalpel finds an occurrence of a recursive function it recovers the beginning of the recursive call and does not inline it. Instead, \prodscalpel extracts the function to a file with the same name of the where the function is implemented. Then, it inserts a calling from the vein to the recursive function. This solution provides a finite interpretation for not inlining mutually recursive functions in the array of over-organ statements. Thus, \prodscalpel can produce a search space for GP with a finite amount of program statements even from recursive functions.

To handle transplantation of organs spread in multiple files, \prodscalpel records the name of the files where the slice is and its location with respect to other slices from the over-organ. 
Then, it records the related statements in an Abstract Syntax Tree (AST), according to their order of appearance in the file. 
This is done to preserve the same structure in the transplanted organ as it appeared in the donor.
Then, \prodscalpel computes the resulting slices in copies of its original files  in the transplantation platform, without breaking the over-organ.

\subsection{Automating Over-organ Reduction and Adaptation}\label{sec:organ_reduction}
\prodscalpel improves the process of over-organ reduction and adaptation by being able to handle over-organs containing multiple files.
It also introduces a layer in the organ that works as an organ-host wrapper. 

In practice, \prodscalpel uses GP, as in previous work~\cite{Barr2015}, to prune one or more program elements within the boundaries of the target organ while keeping the organ still functional and passing on the icebox tests. 
In the wrapper, \prodscalpel~abstracts variable names so that GP can select a type-compatible binding. It selects different combinations of all valid statements, variables and function calls mapped from the organ's vein to initialise an execution environment that the organ expects before executing it. 

\prodscalpel uses GP to search for matching between variables in the organ and the product base, during the over-organ adaptation process. The matches found are inserted in the organ-host wrapper. 
By a mutation operation, a new version of the organ (i.e., a new individual) is created while \prodscalpel makes several changes in the organ-host wrapper and pruning the over-organ. Each such mutation operation is either an $\texttt{INSERT}$, $\texttt{REPLACE}$ and $\texttt{DELETE}$ of code into the individual and the wrapper at the level of statements.

In the end, \prodscalpel synthesises a call to the extracted organ to execute and test it from the wrapper constructed.

\subsection{Automating Multiple Organ Implantation}
\label{sec:Implantation}

Once the organ is adapted to correctly work on the host it can be automatically implanted into the product base. To correctly insert an organ into the tool must identify and handle potential implicit connection points by avoiding the insertion of code duplication into the product base.

\begin{figure*}[t]
	\centering \includegraphics[width=0.8\textwidth]{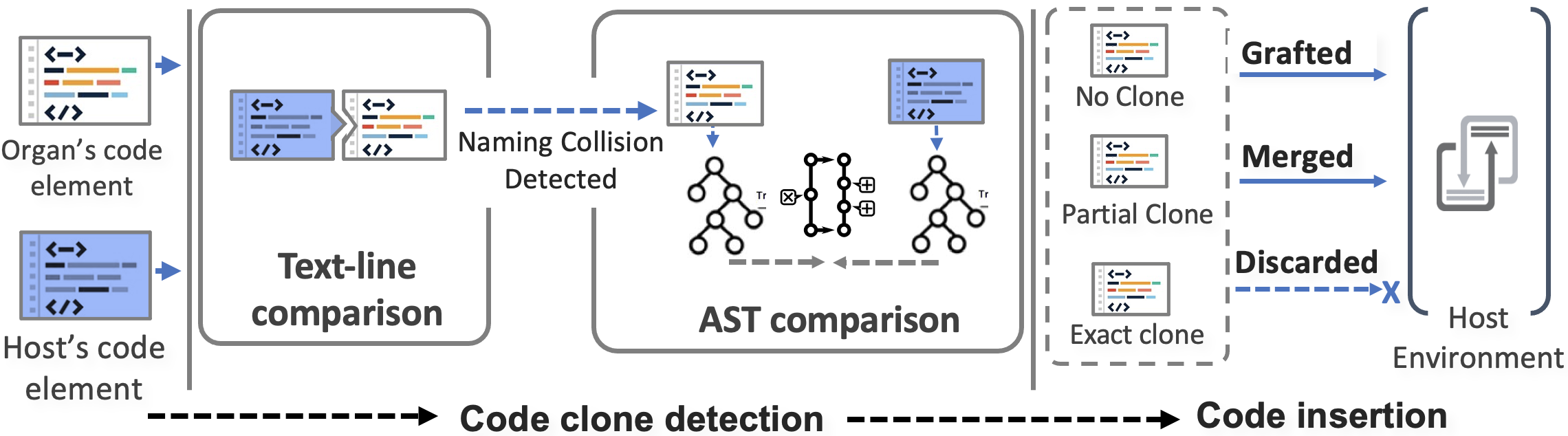}
	\caption{Code clone detector. \textit{It performs two comparison steps: text-line and ASTs comparisons to identify code clones and make decision if a code elements must be graft our not into the target product base environment.}}
	\label{fig:code_clone_analysis}
\end{figure*}

\Cref{fig:code_clone_analysis} illustrates our solution to avoid the organ collision problem. To sum up, the clone detector checks if a specific code element is already present in the beneficiary's environment. To do this, \prodscalpel constructs two lists, one with elements in the organ and other with elements in the host. Then, it checks if there is some element in the target organ which is already presented in a list of code elements transplanted. Once a potential element duplication is identified, \prodscalpel decomposes both organ and host elements in ASTs to detect semantic dependencies between organ implementations previously transplanted, preventing shared elements to be inserted into the host again. Then, the code element is entirely \emph{grafted}, \emph{discarded}, or \emph{merged}. In this last case, \prodscalpel introduces additional line breaks such that potential variances within statements and other structures can be accurately inserted by using sub-abstract tree comparison.

We augmented \prodscalpel with a \emph{code clone detector}, based on NiCad~\cite{Roy2009}. This clone detector finds exact clones over arbitrary program fragments in the organ and host source code by using ASTs. Thus, we exploit the benefits of \emph{Program differencing}~\cite{Kernighan1983} technique and TXL~\cite{Cordy2006} to identify and compare potential syntactic code duplication using text-line and ASTs comparison~\cite{Roy2009}.

%% file: sections/sec06-case_studies.tex
\section{Case Studies} \label{sec:case_studies}
We evaluate \prodscalpel~on two case studies for generating new product variants from real-world systems. This section explains the objectives, research questions, and subject systems. Additionally, it presents the results of our study and discusses the implications. Our corpus and data collected are available at the project webpage~\cite{ProjectWebpage}.

\subsection{Objective and Research Questions}
The objective of this study is to evaluate the proposed approach and tool, thereby demonstrating the potential of automated software transplantation for product line generation. 
We aim to answer the following research questions:

\begin{quote}
\textbf{RQ1.} \emph{How much effort is required to generate products from a product line created using \prodscalpel?}
\end{quote}
Given the complexity of the task at hand, it seems unreasonable to expect the product derivation process to be instantaneous. Still, it will need to be fast enough to be incorporated into a development cycle and demonstrate its advantage over manual effort. To answer this question, we computed the time required, and the number LOCs transferred to productize software.

\begin{quote}
\textbf{RQ2.} \emph{Which features can \prodscalpel~productize?}
\end{quote}
We discuss the characteristics of features that \prodscalpel~can and those it cannot yet handle. Given the inherent complexity of the process, it is important to clearly define its current limitations. 

\subsection{Methodology}

\textbf{Subject systems}. We chose two text editors, \emph{VI} and \emph{VIM}, as product bases.
For donors we chose subjects from two different domains: code analysis software and text editors.
We chose \emph{GNU Cflow}, a call graph extractor from C source code, as a donor.
This is to show that donors can also come from different application domains.
We reuse \emph{VI} and \emph{VIM} as donors, as well as another text editor, \emph{kilo}.
This is to show that donors can come from the same system as the product base, but also different systems within the same application domain.

Our donors are presented in Table~\ref{tab:donors_list}.
We identified the following features as possible desired features in a new editor: $\texttt{output}$ from CFLOW, $\texttt{enableRawMode}$ from kilo, $\texttt{vclear}$ from VI, and $\texttt{spell\_check}$ and $\texttt{search}$ from VIM. Table~\ref{tab:donors_list} presents more details on the subjects.

\textbf{Procedure and execution}. 
Initially, we automatically remove dead-code from both donors and host codebases by using \prodscalpel. 
We also reduced each host to its basic form removing all optional features. Thus, both donors and host are prepared for the transplantation process. 
Given an organs' entry point for each organ in the donor systems provided by us, their target implantation points in the product base, and a set of test suites for each organ, \prodscalpel was used to localise and extract a set of organs from the donor, transform each organ to be compatible with the context of their target sites in the product base and implant the organs in the beneficiary's environment. Each automated organ transplantation process was repeated 20 times, due to the heuristic nature of the over-organ adaptation process. We computed the average number of lines of code transplanted and the average runtimes for the transplantation process.

\textbf{Case study environment}. The runtimes for each transplantation were measured on an Intel Core 3.1 GHz Dual-Core Intel Core i5, with 16 GB memory running MacOS 10.15.4.

\begin{table}[t]
\centering 
	\caption{Donors and hosts corpus for the evaluation: column Features shows the number of features identified. 
	}
	\label{tab:donors_list}
	\begin{tabular}{llrr} \hline
        Subjects &Type  & Size (LOC) & \#Features   \\\hline
	    Kilo     &Donor & 804        & 17      \\
		CFLOW    &Donor & 4,274      & 54      \\
		VI       &Donor & 20,292     & 36      \\
		VIM      &Donor & 839,438    & 176      \\\hline
		VI       &Product Base & 20,223 & 36      \\
		VIM      &Product Base & 737,466 & 117     \\\hline
	\end{tabular}
\end{table}

\subsection{Results and Discussion} \label{sec:result_discussion}

Table~\ref{tab:transplantation_time} shows average runtimes of transplanting each organ into the product base as well as the total number of code lines transplanted to derive each product. At the end of the transplantation process, the postoperative product A has a total of 28k LoC and 40 features while product B has a total of 745k LOC and 121 features. Together, donors provided three feature variants to the product line, approximately 7.8k LOC to product A and 8.1k to product B, including a feature removed and re-transplanted in \emph{VI}.

\begin{table}[t]\centering 

    \caption{Multi-organ transplantation results to generate products A and B. \emph{Columns Trans. Time shows the time (Sys+User) spent on the organ transplant process in which column Sys. correspond to the \prodscalpel's execution time; column User correspond to the user time spent in preoperative stage and  augmenting the host’s regression test suite (regression++). Product A uses a reduced \emph{VI} editor as a product base, while Procut B was derived by transplanting features from the donors into the reduced \emph{VIM} editor.}}
    \label{tab:transplantation_time}
	\begin{center}
	\begin{tabular}{lrrrrrr} \hline
		\multicolumn{1}{c}{Donors} & \multicolumn{3}{c}{Number of}   & &\multicolumn{2}{c}{Trans.time(min)} \\
		\cline{2-4} \cline{6-7}
		 & \multicolumn{1}{c}{LoC}  & \multicolumn{1}{c}{Functions} & \multicolumn{1}{c}{Files} & & \multicolumn{1}{c}{Sys.} & \multicolumn{1}{c}{User}\\\hline
		Kilo        & 963 & 35 & 4 & & 86 & 32\\
		CFLOW       & 4,822 & 37 & 8 &  & 344 &123 \\
		VI          & 1,983 & 5 & 15 &  & 1234 &184\\\hline
		\rowcolor[gray]{.9} Product A  &7,768 & 77 & 27 & & 1,664 & 339 \\\hline
		Kilo        & 981 & 35 & 4 & & 94 & 32\\
		CFLOW       & 4,898 & 37 & 8 &  & 428 & 123\\
		VI          & 2,234 & 5 & 15 &  & 1294 & 184\\\hline
		\rowcolor[gray]{.9} Product B  & 8,113 & 77 & 27 & & 1,890 &532\\\hline
	\end{tabular}
	\end{center}
	
\end{table}

\begin{framed}
\noindent To answer RQ1, on average, \prodscalpel spent 4h31min/1KLoC for transplanting three features into VI, and 4h40min/1KLOC for transplanting the same three features into VIM.
\end{framed}

To answer question \textbf{RQ2}, we also tried to transplant two more features:  $\texttt{spell\_check}$ and $\texttt{search}$,  both containing larger amounts of code, about 104 KLOCs and 153 KLOCs, respectively.

\prodscalpel uses Doxygen~\cite{Doxygen2018}, a source code documentation generator, to generate the call and caller graphs. 
It thus inherits the limitation of generating an imprecise call graph when dealing with function pointers. 
This leads to unneccessary instructions to be copied, while others missing.
Although possible with an additional manual effort, such limitations made our tool unable to automatically extract the larger organs from VIM. 
We can, in the future, turn to \emph{Dynamic analysis}~\cite{Cornelissen2009} and other code manipulating tools to optimise the efficiency of the slicing process, to efficiently identify those statements in the program slice which have influence on the target organ. Investigating more precise techniques is future work.

\begin{framed}
\noindent To answer RQ2, \prodscalpel was able to productise three features from three different systems. However, transplantation of organs from a larger donor exposed some limitations of the technologies used by our tool.
\end{framed}

In summary, these case studies show initial evidence that software transplantation can be successfully used for building product variants automatically by combining features transplanted from real-world systems.
\prodscalpel inherits limitations of the underlying slicing tools, thus its generalisability can be further improved with progress in that domain.
New studies may give more evidence that our approach is also extensible and flexible in other domains, opening perspectives for future work.\looseness=-1

%% file: sections/sec07-experiment.tex
\section{Experimental Evaluation} \label{sec:experiment}


In the previous case studies, we showed the viability of our software transplantation approach, implemented in \prodscalpel, to generate customized products from the existing codebase. Although the validation of the approach is based on empirical evidence, it  is  still important to test its efficiency by comparing it with other tools used to product line migration. Unfortunately, tool support for the reengineering process is limited, in general, they give support for specific activities, such as feature location, refactoring or quality assurance~\cite{Assuncao2017,Kruger2020}. To the best of our knowledge, there is currently no comparable tool that manages to automatically transplant features from distinct donor systems to generate a product line.
Therefore, we compare our approach with current state-of-the-art, namely human effort.
We conducted an experiment that reflects a real-world process of product line migration from existing codebases~\cite{Krueger2002}. In this section, we state our research objectives and describe in detail the experimental setting.\looseness=-1
 Our corpus and data collected are available at  the project webpage~\cite{ProjectWebpage}.
 
\subsection{Goal}
The goal of this experiment is to analyse the effectiveness and efficiency of our approach compared with the manual process of generating a product line from existing systems, performed by SPL  experts. In accordance with the guidelines for reporting software engineering experiments presented in ~\cite{Wohlin2012b}, we have framed our research objectives using the Goal Question Metric (GQM) method suggested by Basili~\cite{Basili1994}. Our goal is to:

\textbf{Analyse} a software transplantation approach to derive product variants 
\textbf{for the purpose of} comparison 
\textbf{with respect to} effectiveness and efficiency 
\textbf{from the point of view of} the researcher
\textbf{in the context of} an SPL project of product line migration from real-world systems.

\subsection{Research Questions}

In order to achieve the stated goal, we defined two quantitative research questions. These are related to the data collected during the period that the experiment was conducted. The questions are described as follows:

\begin{quote}
\textbf{RQ3.} \emph{How successful is \prodscalpel at feature migration when compared with human effort?}
\end{quote}
We would like to understand how well our approach automatically transfers all required code so that the target feature can run in an emergent product line and compare it with the manual process.

\begin{quote}
\textbf{RQ4.} \emph{How much feature migration time can be gained using \prodscalpel compared to the manual process?} 
\end{quote}
With this question, we evaluate the time spent by SPL experts to \emph{extract}, \emph{adapt} and \emph{merge} features to derive new product variants and compare these with the same processes when using \prodscalpel. 

\subsection{Measures}
With the objective to answer our research questions, we defined the measures that must be computed. For each question, one measure was defined. These are described as follows:

\textbf{M1.} For RQ3, the accuracy of our approach is computed by verifying if \prodscalpel successfully migrated new functionalities to a product line by passing all the regression, augmented regression and acceptance test suites. This will check whether or not the output of the transplanted feature is correct with respect to the given test suites. 

\textbf{M2.} For RQ4, we simply report the time that is spent on each activity to transfer the target features. These activities are: \emph{code extraction}, \emph{adaptation}, and \emph{merging}. The time for each of these activities was collected individually.

\subsection{Methodology}

We use two donor systems in our experiment:
\emph{NEATVI}\footnote{https://github.com/aligrudi/neatvi}, text editor extended from VI for editing bidirectional UTF-8 text and \emph{Mytar}\footnote{https://github.com/spektom/mytar}, an archive manager.  
Table~\ref{tab:instrumentation} gives more details about the systems used in this experiment. Having in mind manually inspecting a codebase to transfer a feature to a product line is hard, slow, and tedious~\cite{Mahmood2021}, we chose to select small systems to avoid participants getting tired. These codebases were available to download together with a script to automate setup of the environment.

\begin{table}[ht]\centering \small
	\caption{Details of donors and product base systems used in our studdy.}
	\label{tab:instrumentation}
   \begin{adjustbox}{width=\columnwidth,center}
    	\begin{tabular}{l|lr|lr|lr} \hline
    		\multicolumn{1}{c}{Scenario} & \multicolumn{1}{|c}{Donors} & \multicolumn{1}{c|}{LoC}   & \multicolumn{1}{c}{Target features}       & \multicolumn{1}{c|}{LoC}  & \multicolumn{1}{c}{Host}& \multicolumn{1}{c}{LoC} \\\hline
    		I       & NEATVI & 5,276 & $\texttt{DIR\_INIT}$  & 239&\multirow{2}{*}{Product base}&\multirow{2}{*}{5,285} \\
    		II      & Mytar  & 1,046 & $\texttt{WRITE\_ARCHIVE}$ & 170&& \\\hline
    	\end{tabular}
   \end{adjustbox}
\end{table}

We recruited 20 SPL experts for the experiment that were divided into two different groups.  We chose to allow participants to use their own work environment by avoiding adaptation bias to a strange environment with the use of unknown tools.
Guidelines provided to the participants of \emph{Group A}\footnote{The guideline for perform the transplantation in  scenario I is available at \url{https://rb.gy/vydhbb}} and \emph{Group B}\footnote{The guideline for perform the transplantation in scenario II is available at \url{https://rb.gy/covz62}} consisted of a process description and systems documentation. Additionally, we provided training on re-engineering software systems to SPL to ensure that all participants understood the experiment's objectives.

We used timesheets to record the effort spent on the necessary activities required to transfer features from an existing codebase to a product base, as previously mentioned. In addition to the timesheets, we used two forms to collect information about participants' experience. This data is shown in Table~\ref{tab:participants}. 
We also conducted a post-survey to better understand participants' problems encountered.  

\normalsize\subsection{Experimental design}

We answer our research questions by simulating a real reengineering process where two features must be transferred to a product line built over a product base. The experimental design was inspired by documented real product-line migration scenarios~\cite{Laguna2013, Assuncao2017}. 

In scenario I, we gathered a group of 10 SPL experts (called Group A) where each one of them had to manually re-transplant all portions of code that implement the feature $\texttt{dir\_init}$ to the product base. We removed this feature from the original version of \emph{NEATVI} to generate the product base used in this scenario. In scenario II, another group of 10 SPL experts (called Group B) tried to insert the feature $\texttt{write\_archive}$ from \emph{MYTAR} into the original version of NEATVI used as the product base. Here, we chose not to provide the post-operative product base used in the scenario I. Instead, we provide the original version of NEATVI, already with the feature $\texttt{dir\_init}$, to prevent possible code errors introduced in scenario I. As we analyse both scenarios separately, we believe that this strategy has minimized possible human bias.

\begin{figure}[t]
	\centering \includegraphics[width=8.5cm]{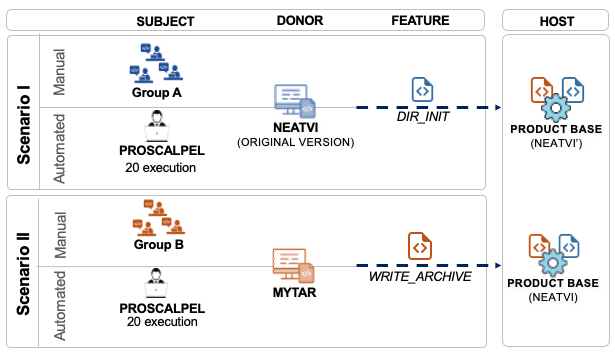}
	\centering 
	\caption{Experimental design: one factor with two treatments applied in two re-engineering scenarios.}
	\label{fig:experiment_design}
\end{figure}

The idea is to use each scenario to represent real scenarios of re-engineering to SPL where system variants came from both similar (scenario I) and distinct codebases (an archive manager providing features to a text editor's product line - scenario II) as Figure~\ref{fig:experiment_design} shows.

In both scenarios participants are given the same inputs as \prodscalpel.
Thus the independent variable in our experiment is whether the feature migration process is automated or not.

In this experiment, the dependent variables are: success of the product line generation process and the payoff for using our approach. That is, to analyze the success of our approach (RQ3), we evaluate how often the transplanted features pass all the provided test cases.
To analyze the performance of the approach (RQ4), we measured the time spent by participants to extract, adapt and merge one feature into a product base in comparison with \prodscalpel's time to complete the same tasks.

\subsection{Pilot Study}

First, we conducted two pilot studies with 6 graduate students. 
We used the pilot study results to determine the amount of time needed to execute our tasks and the suitable size of features. This allowed us to estimate and plan the number of participants we needed for the main study. 
The pilot study also allowed us to assess whether the participants could properly understand the subject systems and the tasks they should perform. 

\subsection{Participants}

After the pilot study, we recruited 20 new participants: 2 undergrads (Un), 9 masters (M), and 9 PhD. (PhD)9. Most of them have more than 5 years of SPL experience and 10 years of software development. The participants are from ten different universities (from U1to U10) and the analysts/developers work in four different companies (C1, C2, C3, C4 and C5). To recruit them, we sent emails to professors from two universities, from different software reuse research groups, to suggest current and ex-members. 

Before the experiment, we asked them to answer an online survey, which we used to collect background data about their experience, mainly in software development and SPL\footnote{\url{https://rb.gy/ant4g8}}. 
We created balanced groups (A and B) of participants for each product line generation scenario, based on their experience. 
Table~\ref{tab:participants} shows the details of the participants involved in the experiment.

\begin{table}[t]
\centering 
	\caption{\textit{Details of participants' expertise (in years) and division into groups. Group A worked on scenario I, transplanting a feature from different versions of the same donor system as the host. Group B worked on scenario II,transplanting a feature from a donor system different to the host one}}
	\label{tab:participants}
	\resizebox{7cm}{!}{%
	\begin{tabular}{clllll} \hline
		Group &Part. &Degree & Inst. & \multicolumn{2}{c}{Exp. (years)}\\ \cline{5-6}
		              &         &  &   & Dev.      &  SPL     \\\hline
		A       &P1  & MSc     & U7   & [1,5)     &  [5,10)  \\
    	        &P2  & Un      & C5   & [10)      &  [1,5)   \\
		        &P3  & PhD     & U4   & [10)      &  [10)    \\
		        &P4  & MSc     & U4   & [10)      &  [1,5)   \\
		        &P5  & MSc     & U4   & [10)      &  [5,10)  \\
		        &P6  & MSc     & U4   & [10)      &  [5,10)  \\
		        &P7  & PhD     & U8   & [10)      &  [10)    \\
		        &P8  & PhD     & U9   & [10)      &  [1,5)   \\
	            &P9  & PhD & U10  & [10)      &  [10)    \\
	            &P10  & PhD     & U4   & [1,5)     &  [1,5)   \\\hline
	            
		B       &P11   & MSc     & C1   & [10)      &  [1,5)   \\
	            &P12   & MSc     & C2   & [1,5)     &  [1,5)   \\
		        &P13   & Un      & C3   & [10)      &  [1,5)   \\
		        &P14   & PhD     & U1   & [10)      &  [10)    \\
		        &P15   & PhD     & U2   & [10)      &  [10)    \\
		        &P16   & PhD     & U3   & [10)      &  [5,10)  \\
		        &P17   & MSc     & U4   & [5,10)    &  [5,10)  \\
		        &P18   & MSc     & C4   & [5,10)    &  [5,10)  \\
		        &P19   & PhD     & U5   & [10)      &  [10)    \\
		        &P20  & MSc     & U6   & [5,10)    &  [1,5)   \\\hline
		        
	\end{tabular}
	}
\end{table}

\subsection{Operation} 

Before the participants receive their tasks, we introduced the experiment with a \emph{tutorial} on reengineering of existing systems into SPL. The tutorial took 30 minutes on average.

We provided the participants with the same input as the one required for \FOUNDRY, namely: feature entry points in the donor, the donor’s source code, and a prepared product base with the target insertion point.
We also supplied the participants with ice-box tests that could be used to guide the search for organ code modifications required to check if it continues to be executable when deployed in the host. Additionally, they received a few-sentence description of each feature in the target system and the system’s documentation with donor and host feature models. 

The direct costs of this experiment are related solely to the time spent by the researcher with the setup of the experiment itself. 
This involved: specifying the respective annotations for the entry point and insertion points of the features, which took approximately 13 minutes of work for scenario I and 17 minutes for scenario II; creating the test cases, necessary to validate the target features, taking approximately 16 minutes of programming activities; preparing the product base, which took approximately 14 minutes; then, approximately 34 minutes were spent creating all documentation of donor systems including the product base feature model.

To extend the emergent product line with a new feature transferred from the correspondent donor system, all participants were required to perform three activities based on the migration of system variants to the target product line~\cite{Krueger2001,Mahmood2021}: feature \emph{extraction}, \emph{adaptation} and \emph{merging}, with descriptions and instructions provided for each task.  

Initially, the participants had to identify and extract all code associated with the feature of interest to a temporary directory. 
Then, each participant had to adapt the extracted feature so it executes correctly in the product base environment, passing all unit tests. In practice, each participant had to change the feature source code to be compatible with the name space and context of its target insertion point in the product base. Finally, it had to insert the feature's code into the product base, and validate its correctness via regression testing.

\subsection{Data collection}

We have provided a task and time registration worksheet. While participants were conducting the assigned tasks, we asked them to take notes of which strategies were being used for each stage of the feature transfer process and why they are performing each specific task. It allowed us to capture strategies and performance data simultaneously. 

We have complemented the above setup with a post-survey. This way we can better understand participants' problems and differences between the manual and automated process in both scenarios. We have triangulated the data generated from the experiment with the responses we obtain from the pre and post-surveys.

To establish the time for feature transplantation using our automated approach, we ran \prodscalpel 20 times, and measure the average time spent on feature migration in each scenario. This average time was compared with the time spent by our participants on the manual re-engineering process.

Based on our pilot study, we set a time limit of 4 hours for each manual and automated process. 
This is to allow for enough time to transfer required amount of LoC while avoiding participants getting bored or tired.

\subsection{Data Analysis}
\label{sec:experiment_data_analysis}
We used 22 pre-existing regression test suites designed by the \emph{NEATVI} developers to assess the success of \prodscalpel and manual transplantations and answer our \textbf{RQ3}. 
However, these were not designed to test \emph{NEATVI} as a product base with new variants and may not be sufficiently rigorous to find regression faults introduced by the re-engineering process. 
To achieve better product line coverage, we manually augmented the host’s regression test suites with additional tests, our augmented regression suites.
Furthermore, we implemented an acceptance test suite for evaluating the transferred feature in both scenario I and II.
We have a total of 30 such tests in scenario I and 33 in scenario II. 
Our test suites provided statement coverage of 72.5\% to the post-operative product line in scenario I and 73.3\% to the post-operative product line in scenario II.

\subsection{Results and Discussion}
\label{sec:evaluation_results}

We summarise our results in Table~\ref{tab:transplantation_results}.  
We report the status of the product base and feature inserted by the participants, the time spent and the number of passing tests for the regression augmented regression and acceptance test suites \footnote{The time measured for the participants is available at \url{rb.gy/ant4g8}}. 
In the first scenario, only one of the participants was not able to finish the process before the timeout. On the other hand, half of the participants were able to finish the process before achieving the timeout in the second scenario and only three of them have been able to insert the target feature without breaking the product base. 

\begin{table*}[t]
\centering 
    \caption{\textbf{Scenario I:} Donor: NEATVI - Product Base: NEATVI without the desired feature. \textit{Experiment results comparing the time of tool over 20 repetitions with the participants:column product line status shows the generated product line status by participants and tool; column \emph{Execution Time} shows the time spent on the feature transplant by the participants and the average time of 20 runs of \prodscalpel. We highlight the execution time of the participant that moqt quickly completed the task. Columns in \emph{Test Suites} show results for each test suite and report statement coverage (\%) for the postoperative host and for the organ. Columns marked with \emph{PASSED} report the number of passing tests.}}
	\label{tab:transplantation_results}
	\begin{tabular}{lcrrrr}\\\hline
		\multicolumn{1}{c}{}       & & & \multicolumn{3}{c}{Test Suites} \\            
		\cline{4-6}  
		 \multicolumn{1}{c}{Participants} & \multicolumn{1}{c}{ Product Line} & \multicolumn{1}{c}{Execution}& \multicolumn{1}{c}{Regre. (59\%)} & \multicolumn{1}{c}{Regre.++ (70.1\%)} & \multicolumn{1}{c}{Accept. (72.5\%)} \\
		\multicolumn{1}{c}{} & \multicolumn{1}{c}{Status} & \multicolumn{1}{c}{Time (minutes)} & \multicolumn{1}{c}{PASSED}  & \multicolumn{1}{c}{PASSED}  & \multicolumn{1}{c}{PASSED} \\\hline

		 P1 &\multicolumn{1}{c}{OK}  &\textbf{82}  & 22/22 &30/30 &3/3 \\
		     P2 &OK      &\textbf{88}  & 22/22  &30/30  &3/3\\
		    P3 &OK     &\textbf{77}  & 22/22 &30/30   &3/3\\
		     P4 &OK      &\cellcolor[gray]{.9}\textbf{68}  & 22/22  &30/30 &3/3\\
		     P5 &OK      &\textbf{81}  & 22/22  &30/30  &3/3\\
		     P6 &Broken & \multicolumn{1}{c}{\textbf{Timeout}} & 0/22     &0/30     &0/3 \\
		     P7 &OK      &\textbf{87}  & 22/22 &30/30   &3/3  \\
		     P8 &OK     &\textbf{83}  & 22/22 &30/30   &3/3 \\
		     P9 &OK      &\textbf{73}  & 22/22 &30/30  &3/3\\
		     P10 &OK     &\textbf{113} & 22/22  &30/30  &3/3\\
		 \hline 
		 \rowcolor[gray]{.9} \textbf{\prodscalpel} &\textbf{OK in 20/20 runs} &\textbf{20} &\textbf{22/22}   &\textbf{30/30}   &\textbf{3/3} \\
		 \hline
	\end{tabular}
    \bigskip
    \bigskip
    \bigskip
    \caption{\textbf{Scenario II:} Donor: Mytar - Product Base: NEATVI. All other columns are the same as in the previous table.}\label{tab:B}
	\begin{tabular}{lcrrrr}\\\hline
	\multicolumn{1}{c}{}       & & & \multicolumn{3}{c}{Test Suites} \\                       
		\cline{4-6}  
		\multicolumn{1}{c}{Participants} & \multicolumn{1}{c}{ Product Line} & \multicolumn{1}{c}{Execution}& \multicolumn{1}{c}{Regre. (70.1\%)} & \multicolumn{1}{c}{Regre.++ (71.9\%)} & \multicolumn{1}{c}{Accept. (73.3\%)} \\
		\multicolumn{1}{c}{} & \multicolumn{1}{c}{Status} & \multicolumn{1}{c}{Time (minutes)} & \multicolumn{1}{c}{PASSED}  & \multicolumn{1}{c}{PASSED}  & \multicolumn{1}{c}{PASSED}  \\\hline
		 P11 &Broken & \multicolumn{1}{c}{\textbf{Timeout}} & 0/22 & 0/33 & 0/2\\
		 P12 &Broken  & \multicolumn{1}{c}{\textbf{Timeout}} & 0/22 & 0/33 & 0/2\\
		 P13 &Error    &\textbf{181} & 0/22 & 0/33 & 0/2\\
		 P14  &Broken &\multicolumn{1}{c}{\textbf{Timeout}} & 0/22 & 0/33 & 0/2\\
		 P15 &Broken  &\multicolumn{1}{c}{\textbf{Timeout}}  & 0/22 & 0/33 & 0/2\\
		 P16 &Error    &\textbf{114} & 0/52 & 0/33 & 0/2\\
		 P17 &OK       &\cellcolor[gray]{.9}\textbf{104} & 22/22 & 33/33 & 2/2\\
		 P18 &OK       &\textbf{194} & 22/22 & 33/33 & 2/2\\
		 P19 &OK       &\textbf{131} & 22/22 & 33/33 & 2/2\\
		 P20 &Broken & \multicolumn{1}{c}{\textbf{Timeout}} & 0/22 & 0/33 & 0/2\\
		 \hline 
		 \rowcolor[gray]{.9} \textbf{\prodscalpel} &\textbf{OK in 19/20 runs} &\textbf{27} & \textbf{22/22} &\textbf{33/33}  &\textbf{2/2} \\\hline
	\end{tabular}
\end{table*}

For each scenario, we also report the number of \prodscalpel runs in which the product derived passed all test cases. For each scenario, we repeat each run 20 times. The success rate was retained for both scenarios I and II, where only one run timed out and the product line generated passed all tests from all test suites.

The results show success rate was retained in both scenario I and II, where we lost only one successful run in the timeout and all products derived passed all tests from all test suites.  
With regards to the manual process, nine participants have successfully transplanted the target feature in scenario I. 
In scenario II, seven of ten participants broke the product base when trying to transplant the target feature, while half of the participants were not able to transplant the feature within the timeout.

\begin{framed}
\noindent \textbf{RQ3:} Our results show that \prodscalpel outperformed participants in both scenarios with only one run timing out.  
Eight of twenty participants (considering both scenarios) were not able to transplant the target features without breaking tests.
\end{framed}

\begin{figure}[t]
	 \includegraphics[width=8.6cm]{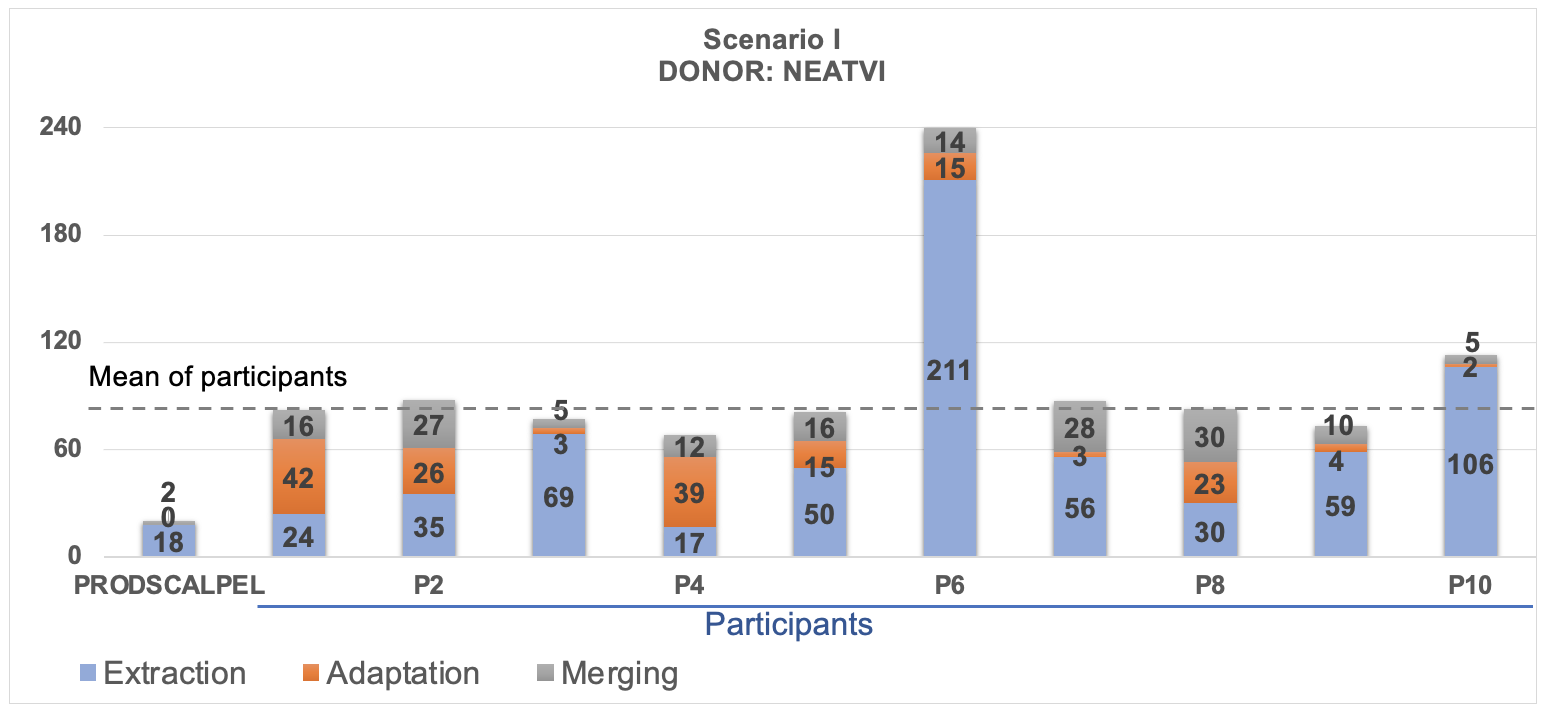}
	\label{fig:experiment_result_time_I}

	 \includegraphics[width=8.6cm]{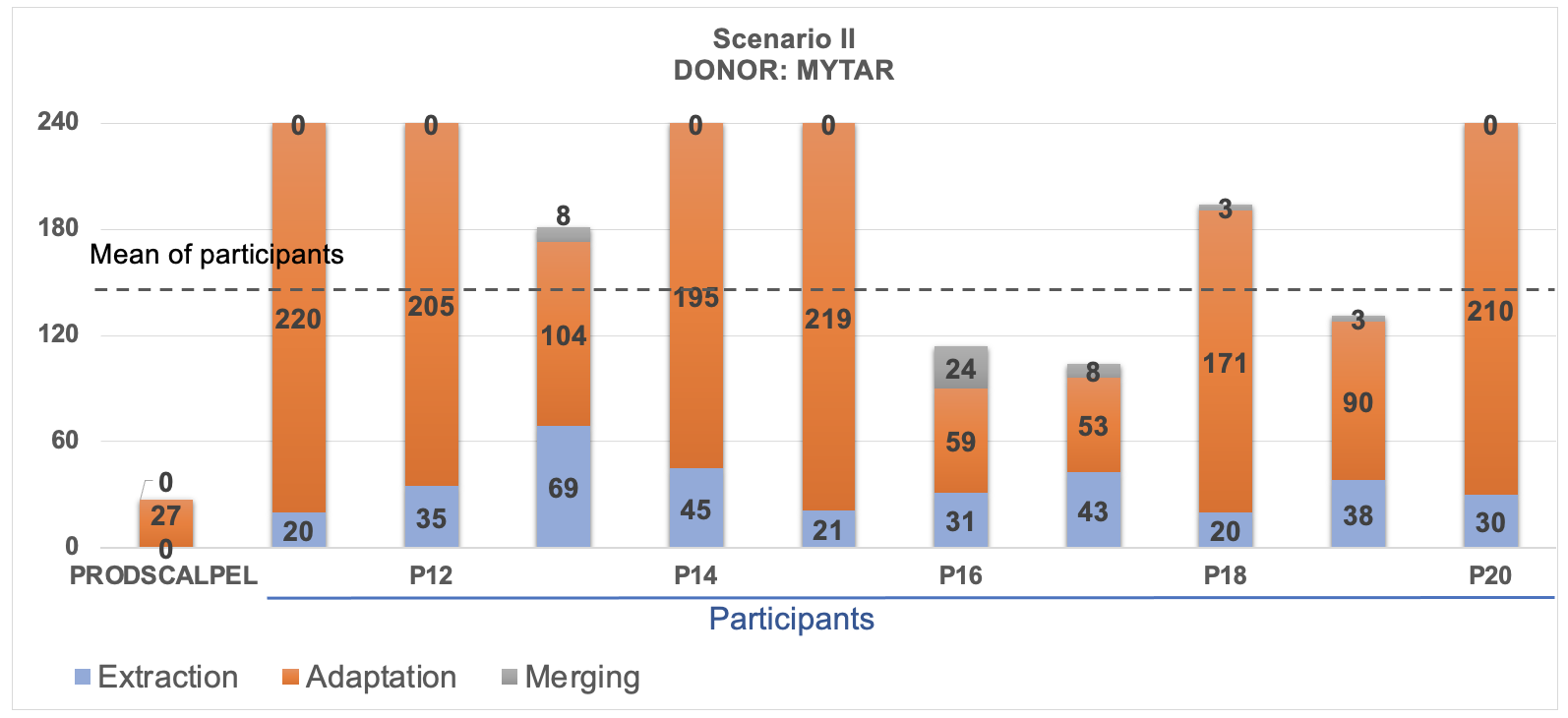}
	 \caption{Time (in minutes) spent by participants and \prodscalpel on performing the three stages of SPL reengineering: feature \emph{extraction}, \emph{adaption} and \emph{merging}. The graph highlights the average time spent by participants who successfully generated new products.} 
	\label{fig:experiment_result_time_II}
\end{figure}

As stated in the definition of measure \textbf{M2} and to answer \textbf{RQ4}, we evaluate the payoff of \prodscalpel. Figure~\ref{fig:experiment_result_time_II} graphically shows the time spent on each activity performed in the re-engineering to SPL process. In summary, Group A transferred the target feature from \emph{NETVI} to the product base in 1h24 minutes on average. \prodscalpel turned out to be quicker, successfully transplanting this feature in all 20 trials, taking an average of 20 minutes.

Most of the participants from Group B had not completed the feature migration process from \emph{Mytar}  within the 4 hours time limit. Considering the participants that were able to finish the process (i.e., participants \emph{P17}, \emph{P18} and \emph{P19}) successfully (all tests passed) they spent an average of 2h23 minutes while \prodscalpel was able to complete this task in 19 of 20 trials in the timeout, taking 27 minutes on average.

By considering the time spent in both scenarios, the tool accomplished the product line generation process 4.8 times faster than the mean time taken by participants who were able to finish the experiment within the timeout.

\begin{figure}[t]
	\centering \includegraphics[width=8.5cm]{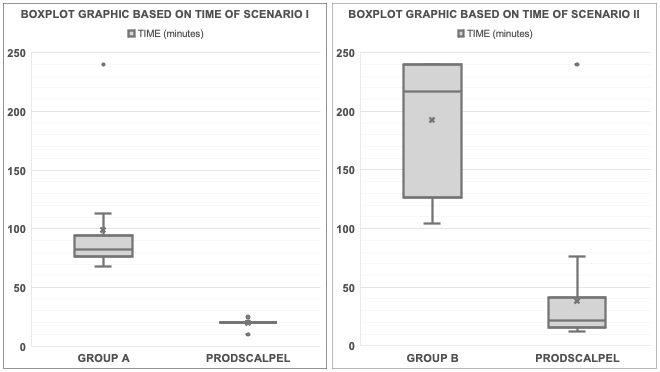}
	\centering 
	\caption{Time results grouping automated and manual in both scenarios.  Scenario I: \emph{NEATVI} - Product Base; Scenario II: \emph{Mytar} - Product base.}
	\label{fig:experiment_result_boxplot}
\end{figure}

\textbf{Statistical analysis of performance.} 
To establish statistical significance of our results we first establish statistical distribution of runtimes for each scenario through the Shapiro-Wilk~\cite{SHAPIRO1965} test. 
Next, we use ANOVA~\cite{Gelman2005} and Pairwise Student’s t-test analysis to test the hypothesis that \prodscalpel's runtimes are statistically lower when compared with the times required by the participants to complete the same task (p-value $<$ 2e-16). We rejected the null hypothesis for all pairs. Figure~\ref{fig:experiment_result_boxplot} graphically shows the time results for our two groups in comparison with \prodscalpel's performance. In scenario I, the preliminary information provided by the box plots indicates that all samples are normally distributed (W = 0.70445, p-value = 3.129e-05).

In scenario II, the normality test result showed a normal distribution with a W = 0.69378, p-value = 1.715e-06. Thus, we used ANOVA to hypothesis testing and Pairwise t-Student. Based on the ANOVA test and Pairwise t-Student, we rejected the null hypothesis (p-value $<$ 2e-16) that the distribution of the population is homogeneous.

We can conclude that \prodscalpel reduces developer effort to transfer features to a product line in both scenarios. For both simulation scenarios, there is a significant effect size between the tasks performed in a manual way and using \prodscalpel. 
Our participants had similar performance times in scenario I, with the exception of participant \emph{P6}. On the other hand, most of the participants of scenario II do not terminate the experiment before the 4-hour timeout. This last one is qualitatively explained by the participants in the post-survey where they mentioned it is hard to adapt a feature from one codebase to run in a different codebase. 
 
\begin{framed}
\noindent \textbf{RQ4} The results confirmed that \prodscalpel outperformed participants transplanting two features 4.8 times faster, on average, compared to participants who finished the process within timeout.
\end{framed}

\section{Threats to Validity}

In this section we present threats to validity of our studies.

\textbf{External Validity.} The relatively small number and diversity of systems used for derive new products pose an external threat to validity. For our user study we applied our results to small programs due to the boundaries of an in-lab study; our results may not generalize to larger programs in the wild. We tried to mitigate it by constructing possible real-world scenarios, i.e., reuse of features from unrelated codebases and variations of the same real-world systems. Additionally, given that our approach was helpful even for small programs, we argue that it is likely helpful for larger systems as it is nearly impossible to incorporate new variants to a product line without a large understanding of the donor systems specifications~\cite{Assuncao2017} and without considerable adaption effort.

\textbf{Internal Validity.} Due to time expensive nature of a human study, we had only 20 participants. We tried to mitigate this issue by selecting participants with considerable experience in SPL projects. The other threat to the validity is the system size; small features are used in this experiment. We assume that inspecting the code to transfer a feature to a product line is hard, which has been confirmed in our post-study survey. We conducted a pilot study to mitigate threats arising from such issues.
Even with a limited execution time, we were able to transplant features from donors of significant size, considering the number of code lines of the subject systems, as shown in Table~\ref{tab:instrumentation}. 
Most participants were also able to complete the given tasks within the time limit.
We also use testing as means of validating our approach, which cannot provide a formal proof of correctness.  We used extensive testing to mitigate this risk. Moreover, testing is a standard approach to code evaluation in real-world scenarios due to its high scalability.

%% file: sections/sec05-automated_spl_reengineering.tex
\section{Automated support for product line re-engeneering} \label{sec:automated_spl_reengineering}

Automated software transplantation, as presented in \FOUNDRY, can provide a set of opportunities to improve and automate the product line re-engineering process. This alone might justify the effort to explore research opportunities in software transplantation for SPLE. Here we elaborate on and discuss such opportunities.

\subsection{ Automating re-engineering of existing systems into SPL} Principally \FOUNDRY is proposed as a new way of re-engineering existing systems into SPL. 
Companies can use \FOUNDRY just for the initial conversion of an existing codebase into SPL. In this mode, \FOUNDRY first extracts over-organs from the donor systems, then, produces a product base, created from an existing system by removing all unwanted features to the product line. Then, adapts and implants the organs into this product base, identifying and removing cross-over-organ redundancies. The resulting organs become the SPL's shared set of features. \FOUNDRY can itself be used as a variability mechanism to construct new products by implanting features into a product base. Even with this one-off application of \FOUNDRY, a company can also use \FOUNDRY to surround the transplanted feature with  a conventional variability mechanism. For existing SPL codebase, such as the one it created. \FOUNDRY's variability mechanism insertion also supports surrounding implanted organs with feature flags or preprocessor directives, which permit enabling and disabling of features, to facilitate its integration into an existing SPL codebase that uses them.

\subsection{Automating clone-and-own technique} One compelling research avenue lies in utilizing \FOUNDRY to automate the application of the \emph{clone-and-own} technique~\cite{Dubinsky2013, Fischer2015}. The \emph{clone-and-own} approach entails duplicating existing software artifacts and independently modifying them to accommodate various product variants. However, managing consistency and synchronization between shared features across these variants presents significant challenges. 

\FOUNDRY can automate \emph{clone-and-own}, especially the task of synchronising changes to a feature shared across two products created by clone-and-own. For example, consider the case where the copy of a shared feature in one of the two products is patched to fix a bug.  With \FOUNDRY, the fixed version of the feature can be transplanted onto the unpatched copy, synchronizing the changes and eliminating the need for manual intervention.

\subsection{Automating Reactive Product Line Adoption Process}

\FOUNDRY, with its automated transplantation capabilities, represents an opportunity to enhance and streamline the reactive product line adoption process. By leveraging \FOUNDRY, organizations can initially generate product variants based on specific demands or requirements. 

As the demand for specific products or features increases, \FOUNDRY can be employed to automate the generation of additional product variants from the transplant of organs extracted from specialized products, effectively evolving it into a product line. 

The use of \FOUNDRY in the reactive product line adoption process offers several benefits. Firstly, it provides a systematic and automated approach to product line expansion, ensuring consistent and controlled feature integration. Secondly, by automating the transplantation of features, \FOUNDRY minimizes the risk of introducing errors or inconsistencies during the adoption process. Furthermore, \FOUNDRY can ve used to guide a modular and incremental growth of a product line. It allows organizations to incrementally introduce new features or product variants in response to market demands. 

Further research can explore and refine the automation capabilities of \FOUNDRY in the context of reactive product line adoption. This includes investigating techniques for efficient feature identification and prioritization, optimizing the automation process for generating product variants, and developing strategies for effectively managing the evolving product line.


\subsection{Automating a symbiotic SPL} \FOUNDRY introduces a novel approach to SPL called ``symbiotic SPL", enabling a symbiotic relationship between the donor codebase and the ongoing reorganization of the SPL. In this mode, the donor codebase remains oblivious to the parallel SPL reengineering process, allowing for independent improvements to be made in both the donor and host codebases.

A key feature of \FOUNDRY is the ability to capture and incorporate improvements from the donor codebase. Periodically, \prodscalpel refreshes its set of features by re-transplanting them into the transplantation platform and subsequently into the host products. This symbiotic approach allows for the integration of updates and enhancements into the SPL, without disrupting the ongoing reorganization efforts.

For example, \prodscalpel can be utilized to create lightweight and specialized text editors from the \emph{VIM} project. By extracting and transplanting specific features from \emph{VIM}, \prodscalpel can generate new text editors tailored to specific user needs or requirements. This symbiotic SPL approach enables the creation of specialized products while leveraging the ongoing development and improvements in the donor codebase.

The symbiotic SPL paradigm can offer several advantages. Firstly, it allows for a more dynamic and responsive development process, as improvements in the donor codebase can be quickly assimilated into the host products. Such strategy can ensure that the SPL remains up-to-date with the latest enhancements, bug fixes, and optimizations.

Secondly, the symbiotic approach enhances the modularity and maintainability of the SPL. By encapsulating features within transplantable over-organs, the donor codebase can evolve independently, enabling the introduction of new features or improvements without affecting the host products. This modularity promotes code reuse can facilitate maintenance, and simplifies the management of feature variations.

The symbiotic SPL facilitated by \FOUNDRY opens up new possibilities for SPLE. It combines the advantages of independent development in the donor codebase with the systematic and automated feature integration provided by SPL.

\subsection{Supporting Product Line Maintenance and evolution of SPL}
The maintenance of assets and products is a challenging task and an inherent characteristic of a product line. When we include organs as assets in a product line the process becomes even more challenging. In fact, by trying to maintain an over-organ individually in the platform, a developer might introduce errors into the product line or in products derived from it, since their organs eventually share elements —such as variables and functions—with the maintained organ. A solution would be to re-implant the changed organ. Nevertheless, a crucial problem here is how to re-implant the organ changed to match a different version of it already transplanted in the target product. 

We envision two ways to avoid this problem and \emph{maintain} a created product line. Firstly, one can re-transplant the features if the original source codebase changes. Secondly, one can maintain the extracted over-organs, and re-run the adaptation and implantation stages, as need be.

In both scenarios of product line maintenance, \FOUNDRY can be set up to evolve a changed organ with feature flags. Surrounding an organ with feature toggles just before implantation is an interesting alternative for continuous deployment of organs. It thus separates inserted code and makes it easier to maintain.
To support continuous deployment, the implantation process in \FOUNDRY can be configured to involve each organ with feature flags to connect new, unreleased code to production. Once a transplanted organ is ready for production, developers can turn off the flag and reveal the new organ or its changes to users.

\subsection{Providing a Variability Mechanism through Organ Transplantation} 

\FOUNDRY introduces a powerful variability mechanism that leverages the transplantation of organs into product bases. This mechanism offers developers the flexibility to include or exclude specific features, enabling the instantiation of different products at any given moment by simply transplanting the corresponding organs into the target product.

This approach addresses the challenges associated with maintaining and evolving a product line that consists of a vast number of individual products. Instead of managing and maintaining a multitude of product configurations, \FOUNDRY streamlines the process by allowing developers to selectively incorporate desired features through organ transplantation. This variability mechanism eliminates the need for extensive feature toggle management and reduces technical debt related to the presence of unremoved feature toggle.


%% file: sections/sec08-related_work.tex
\section{Related Work} \label{sec:related_work}

In this section, we present an overview of existing research related to our work. We position our work within the existing body of knowledge in areas of re-engineering of systems into SPL, clone-and-own, variability in SPL and software transplantation. 

\textbf{Re-engineering of Software Systems into SPL}. Diverse academic proposals and industrial experience reports addressing re-engineering of legacy systems into SPL are present in the literature~\cite{Assuncao2017}. 
However, this number decreases considerably when we are interested in proposals that automate the lifecycle of the re-engineering process~\cite{Kruger2020}. 
 
Martinez et al.~\cite{Martinez2015} introduced \emph{But4Reuse}, a generic and extensible open source tool to facilitate extractive SPL adoption. But4Reuse is a tool that aims to extract SPLs from legacy systems by identifying a set of reusable assets and representing them in a modular way. The tool uses a variety of program analysis techniques, including clone detection and feature location, to identify commonalities and variabilities in the code. Once the SPL is extracted, But4Reuse generates a set of variability models, which can be used to configure the SPL for different product variants.
In contrast, \FOUNDRY does not assume an existing set of product variants. 
SPL can be created from a single codebase, and only requires feature entry point annotation, and a set of tests.
The needed code is automatically extracted using slicing, and modified to run in the given product base via automated over-organ adaptation.

\emph{IsiSPL}~\cite{hlad2021} is a reactive approach~\cite{Krueger2001} to SPL adoption. 
IsiSPL automates the integration of new products into an existing SPL and thus generation of a new SPL with the new features.
In particular, whenever a new product is added, a list of all features needs to be provided.
IsiSPL then analyses SPL to only insert new features, annotating them with conditional directives. However, with large number of products inserted over time, the list of conditional directives will grow, hindering code comprehension, maintenance, and ease of derivation of new products. In \FOUNDRY the use of such directives to handle variability is not a mandatory condition to use it. The developer is free to automatically introduce conditional directives in the evolved product line. That is they can also generate a product without surrounding the feature's code with additional directives, simply by transplanting organs from the transplantation platform. 

\textbf{Clone-and-own}. \FOUNDRY can be exploited as an automated alternative to clone-and-own, where, instead of creating a product line, products are cloned and amended, based on demand.
Although there exists automated support for feature detection using code clone detection, its adaptation for reuse still requires manual work~\cite{Yoshimura2006, Kastner2014}. 

Fischer et al.~\cite{Fischer2015}, for instance, present \emph{ECCO} to enhance clone-and-own. 
The tool finds the proper software artifacts to reuse and then provides guidance during manual adaptation phase, by hinting which software artifacts may need to be migrated and adapted. 
Moreover, ECCO requires that the features' source code must be extracted from the same family of products, which limits its ability to reuse assets. 
In contrast, \FOUNDRY stores over-organs that can be automatically adapted to different product bases. These do not need to come from the same family of products as the product base. Once extracted, features implemented by stored over-organs can be adapted, and implanted into a product base in a fully automated way.

\textbf{Variability in SPL.} The capacity of providing variability in a software development process is a key aspect of modern software development, enabling software products to be customized and adapted to meet the needs and requirements of different stakeholders. 

Several works have already identified the frequent use of the \emph{variability mechanisms}, like preprocessor directives~\cite{Liebig2010} and feature flags~\cite{ Kruger2018, Meinicke2020}, as strategies for allowing the inclusion or exclusion of specific code blocks or features in the product line at compile time or runtime. Both are annotation-based implementation techniques for software product lines require explicit annotations often scattered across multiple code units (e.g., preprocessor annotations such as $\texttt{\#IFDEFs}$)~\cite{Apel2013}. 
These annotations establish a mapping of portions of code to features defined in a variability model. This mapping serves as input to the configurator tools, which then uses the information to select and configure the appropriate features for a given software product.

Despite its error-proneness and low abstraction level, the preprocessor directives are still widely used in present-day software projects to implement variability, maintain, evolve, reuse, or re-engineer a software system~\cite{Kruger2018}. Liegib et al.~\cite{Liebig2010} presents a study of 40 SPL that use preprocessor-based variability mechanisms. The study analyzes the variability mechanisms used in the product lines and the impact of these mechanisms on the codebase, in terms of code size, complexity, and maintainability. Christian Kästner et al.~\cite{Kastner2008} also discuss the concept of variability based on preprocessor directives in software product lines and how it affects the granularity of features. They present a case study of the Linux kernel to illustrate how the different levels of granularity in feature implementation can affect product line evolution and maintenance. 

Other authors, such as, Jezequel et al.~\cite{Jezequel2022} present a case study of how feature models and feature toggles can be used in practice to manage variability in software systems. The authors describe how they used feature models to capture the commonalities and variabilities of a software product line and then translated them into feature toggles that could be used to enable or disable specific features at runtime.

Although useful, these traditional variability mechanisms have limitations~\cite{Bachmann2005, Liebig2010}. Preprocessor directives can lead to code bloat and reduced maintainability, while feature flags can add complexity and overhead to the codebase. Rahman et al.~\cite{Rahman2016} analyzed feature flag usage in the open-source code base behind Google Chrome, finding that feature flags are heavily used but often long-lived, resulting in additional maintenance and technical debt. Meinicke et al.~\cite{Meinicke2020} also discovered that despite the temporary nature of feature toggles and developers' initial intention to remove them, they tend to remain in the codebase unless compelled by policy or technical measures. 

Particularity, building an SPL , where the number of options can grow considerably ~\cite{Lotufo2010,Xu2015}, the use of feature flags and/or preprocessor directives can lead to a large codebase in the emergent product line~\cite{Meinicke2020}. Foundry can be an interesting alternative to those traditional variability techniques. In contrast to the existing approaches where lots of annotations that are permanently added and their number increases over time, our solution requires only an annotation of the feature entry point and its insertion point in the product base. Thus, Foundry's approach has the potential to reduce code complexity and increased readability, as such any extra annotations are not required. Furthermore, Foundry keeps all reusable features of a product line (so-called organs) functional and physically separate, integrating them into the product base only when required for composing a new product.

Overall, ST technique can potentially offer several advantages over the use of traditional feature toggles, including simplified maintenance, reduced code complexity, and reduced risk of conflicts. Furthermore, it can be used in conjunction with existing variability mechanisms like preprocessor directives and feature flags, allowing developers to take advantage of the benefits of both approaches such as the possibility of enabling or disabling organs at runtime or compile-time. 

\textbf{Software Transplantation}. As far as the literature on automated software transplantation is concerned, Petke et al.~\cite{Petke2014,Petke2018} were the first to transplant snipets of code from various versions of the same system to improve its performance, using genetic improvement~\cite{Petke18}. One year later, Barr et al. successfully transplanted a feature from one program into another~\cite{Barr2015}. 

Stelios Sidiroglou-Douskos et al.~\cite{Sidiroglou2017} proposed another tool, CodeCarbonCopy (CCC), which can also transplant code automatically. CCC is a code-transferring tool from a donor into a host codebase. It implements a static analysis that identifies and removes irrelevant functionalists that are irrelevant to the host system. It has performed well in eight code transfers across six applications. However, the code redundancy problem still persists.  CCC is thus unable to handle multiple feature transplantation. Foundry on other hand, addresses this significant problem by exploiting code clone detection. Additionally, CCC inherits the limitations of its static analysis technique~\cite{Bessey2010} to identify the feature codes. It typically only looks at the code in isolation and does not consider the broader context in which the code is used, such as external dependencies or interactions with other features~\cite{Bessey2010, Ilyas2016}.


More recently, Liu et al.~\cite{Liu2018} introduced a method to transplant code from open source software. The validation results indicated that their method can substantially reduce the workload of programmers and is applicable to real-world open-source software. However their idea is also based on program slicing, it does not support the transplant of multiple organs to re-engineering of systems into SPL. Furthermore, they still not provide any tool that support their method.

In contract to those works, we are the first to use automated software transplantation to automate SPL engineering tasks. Our approach and tool can transplant multiple organs to compose an SPL from existing donor systems. In the process, we have solved several issues, previously not considered in the software transplantation literature, such as code redundancy, organ dependence and feature interactions.

%% file: sections/sec09-conclusion.tex
\section{Conclusions and Future Work} \label{sec:conclusion} 

In this work, we propose an approach, \FOUNDRY, and a tool, \prodscalpel, that use software transplantation to speed conversion to and maintenance of SPL.
Both approach and the tool have been validated through case studies. 
We generated two products through the transplantation of features extracted from three real-world systems into two different product bases. 
Moreover, we performed an experiment with SPL experts to compare our approach with manual effort.
We showed significant time effort improvements when using \prodscalpel. 
The tool accomplished product line generation process by migrating two features 4.8 times faster than the mean time spent by participants who were able to finish the experiment within the timeout.

We argue that the migration to SPL transplantation-based in contrast to a configuration-based software product line makes it easier to use in practice.
Our approach improves SPL maintainability by physically separating features from the product base.  
\FOUNDRY can be used both for \emph{extractive} or \emph{reactive} product line migration, as well as as a systematic Clone\&own strategy to specialize existing products.
\FOUNDRY avoids code duplication of feature implementations, while preserving feature behavior. 
It can automatically propagate feature changes. 
That is, it provides solutions for problems often cited in re-engineering of systems into SPL literature.

Our evaluation studies provide initial evidence to support the claim SPLE 
using software transplantation, is a feasible and, indeed, promising direction for SPL research and practice.
However, more studies are needed to provide more evidence for generalisability of our approach, and to investigate its applicability in an industrial context.

%% file: main.bbl

\begin{thebibliography}{56}


\ifx \showCODEN    \undefined \def \showCODEN     #1{\unskip}     \fi
\ifx \showDOI      \undefined \def \showDOI       #1{#1}\fi
\ifx \showISBNx    \undefined \def \showISBNx     #1{\unskip}     \fi
\ifx \showISBNxiii \undefined \def \showISBNxiii  #1{\unskip}     \fi
\ifx \showISSN     \undefined \def \showISSN      #1{\unskip}     \fi
\ifx \showLCCN     \undefined \def \showLCCN      #1{\unskip}     \fi
\ifx \shownote     \undefined \def \shownote      #1{#1}          \fi
\ifx \showarticletitle \undefined \def \showarticletitle #1{#1}   \fi
\ifx \showURL      \undefined \def \showURL       {\relax}        \fi
\providecommand\bibfield[2]{#2}
\providecommand\bibinfo[2]{#2}
\providecommand\natexlab[1]{#1}
\providecommand\showeprint[2][]{arXiv:#2}

\bibitem[Apel et~al\mbox{.}(2013)]%
        {Apel2013}
\bibfield{author}{\bibinfo{person}{Sven Apel}, \bibinfo{person}{Don Batory},
  \bibinfo{person}{Christian Kstner}, {and} \bibinfo{person}{Gunter Saake}.}
  \bibinfo{year}{2013}\natexlab{}.
\newblock \bibinfo{booktitle}{\emph{Feature-Oriented Software Product Lines:
  Concepts and Implementation}}.
\newblock \bibinfo{publisher}{Springer Publishing Company, Incorporated}.
\newblock
\showISBNx{3642375200, 9783642375200}


\bibitem[Assunç{\~a}o et~al\mbox{.}(2017)]%
        {Assuncao2017}
\bibfield{author}{\bibinfo{person}{Wesley K.~G. Assunç{\~a}o},
  \bibinfo{person}{Roberto~Erick Lopez-Herrejon}, \bibinfo{person}{Lukas
  Linsbauer}, \bibinfo{person}{Silvia~R. Vergilio}, {and}
  \bibinfo{person}{Alexander Egyed}.} \bibinfo{year}{2017}\natexlab{}.
\newblock \showarticletitle{Reengineering legacy applications into software
  product lines: a systematic mapping}.
\newblock \bibinfo{journal}{\emph{Empirical Software Engineering}}
  \bibinfo{volume}{22} (\bibinfo{year}{2017}), \bibinfo{pages}{2972--3016}.
\newblock


\bibitem[Bachmann and Clements(2005)]%
        {Bachmann2005}
\bibfield{author}{\bibinfo{person}{Felix Bachmann} {and} \bibinfo{person}{Paul
  Clements}.} \bibinfo{year}{2005}\natexlab{}.
\newblock \bibinfo{booktitle}{\emph{Variability in Software Product Lines}}.
\newblock \bibinfo{type}{{T}echnical {R}eport} CMU/SEI-2005-TR-012.
  \bibinfo{institution}{Software Engineering Institute, Carnegie Mellon
  University}, \bibinfo{address}{Pittsburgh, PA}.
\newblock


\bibitem[Barr et~al\mbox{.}(2015)]%
        {Barr2015}
\bibfield{author}{\bibinfo{person}{Earl~T. Barr}, \bibinfo{person}{Mark
  Harman}, \bibinfo{person}{Yue Jia}, \bibinfo{person}{Alexandru Marginean},
  {and} \bibinfo{person}{Justyna Petke}.} \bibinfo{year}{2015}\natexlab{}.
\newblock \showarticletitle{Automated Software Transplantation}. In
  \bibinfo{booktitle}{\emph{Proceedings of the 2015 International Symposium on
  Software Testing and Analysis}} (Baltimore, MD, USA)
  \emph{(\bibinfo{series}{ISSTA 2015})}. \bibinfo{publisher}{ACM},
  \bibinfo{address}{New York, NY, USA}, \bibinfo{pages}{257--269}.
\newblock


\bibitem[Basili et~al\mbox{.}(1994)]%
        {Basili1994}
\bibfield{author}{\bibinfo{person}{Victor~R. Basili},
  \bibinfo{person}{Gianluigi Caldiera}, {and} \bibinfo{person}{H.~Dieter
  Rombach}.} \bibinfo{year}{1994}\natexlab{}.
\newblock \showarticletitle{The Goal Question Metric Approach}.
\newblock In \bibinfo{booktitle}{\emph{Encyclopedia of Software Engineering}}.
  \bibinfo{publisher}{Wiley}.
\newblock


\bibitem[Bastos et~al\mbox{.}(2015)]%
        {Bastos2015}
\bibfield{author}{\bibinfo{person}{Jonatas~Ferreira Bastos},
  \bibinfo{person}{Paulo~Anselmo da Mota Silveira~Neto},
  \bibinfo{person}{Eduardo~Santana de Almeida}, {and}
  \bibinfo{person}{Silvio~Romero de Lemos~Meira}.}
  \bibinfo{year}{2015}\natexlab{}.
\newblock \showarticletitle{Software Product Lines Adoption: An Industrial Case
  Study (Keynote)}. In \bibinfo{booktitle}{\emph{Proceedings of the Third
  International Workshop on Conducting Empirical Studies in Industry}}
  (Florence, Italy) \emph{(\bibinfo{series}{CESI '15})}.
  \bibinfo{publisher}{IEEE Press}, \bibinfo{address}{Piscataway, NJ, USA},
  \bibinfo{pages}{35--42}.
\newblock


\bibitem[Berger et~al\mbox{.}(2013)]%
        {Berger2013}
\bibfield{author}{\bibinfo{person}{Thorsten Berger}, \bibinfo{person}{Ralf
  Rublack}, \bibinfo{person}{Divya Nair}, \bibinfo{person}{Joanne~M. Atlee},
  \bibinfo{person}{Martin Becker}, \bibinfo{person}{Krzysztof Czarnecki}, {and}
  \bibinfo{person}{Andrzej k{a}sowski}.} \bibinfo{year}{2013}\natexlab{}.
\newblock \showarticletitle{A Survey of Variability Modeling in Industrial
  Practice}. In \bibinfo{booktitle}{\emph{Proceedings of the Seventh
  International Workshop on Variability Modelling of Software-intensive
  Systems}} (Pisa, Italy) \emph{(\bibinfo{series}{VaMoS '13})}.
  \bibinfo{publisher}{ACM}, \bibinfo{address}{New York, NY, USA}, Article
  \bibinfo{articleno}{7}, \bibinfo{numpages}{8}~pages.
\newblock


\bibitem[Bessey et~al\mbox{.}(2010)]%
        {Bessey2010}
\bibfield{author}{\bibinfo{person}{Al Bessey}, \bibinfo{person}{Ken Block},
  \bibinfo{person}{Ben Chelf}, \bibinfo{person}{Andy Chou},
  \bibinfo{person}{Bryan Fulton}, \bibinfo{person}{Seth Hallem},
  \bibinfo{person}{Charles Henri-Gros}, \bibinfo{person}{Asya Kamsky},
  \bibinfo{person}{Scott McPeak}, {and} \bibinfo{person}{Dawson Engler}.}
  \bibinfo{year}{2010}\natexlab{}.
\newblock \showarticletitle{A Few Billion Lines of Code Later: Using Static
  Analysis to Find Bugs in the Real World}.
\newblock \bibinfo{journal}{\emph{Commun. ACM}} \bibinfo{volume}{53},
  \bibinfo{number}{2} (\bibinfo{date}{feb} \bibinfo{year}{2010}),
  \bibinfo{pages}{66–75}.
\newblock


\bibitem[{Bockle} et~al\mbox{.}(2004)]%
        {Bockle2004}
\bibfield{author}{\bibinfo{person}{G. {Bockle}}, \bibinfo{person}{P.
  {Clements}}, \bibinfo{person}{J.~D. {McGregor}}, \bibinfo{person}{D.
  {Muthig}}, {and} \bibinfo{person}{K. {Schmid}}.}
  \bibinfo{year}{2004}\natexlab{}.
\newblock \showarticletitle{Calculating ROI for software product lines}.
\newblock \bibinfo{journal}{\emph{IEEE Software}} \bibinfo{volume}{21},
  \bibinfo{number}{3} (\bibinfo{year}{2004}), \bibinfo{pages}{23--31}.
\newblock


\bibitem[Breivold et~al\mbox{.}(2008)]%
        {Breivold2008}
\bibfield{author}{\bibinfo{person}{H.P. Breivold}, \bibinfo{person}{S.
  Larsson}, {and} \bibinfo{person}{R. Land}.} \bibinfo{year}{2008}\natexlab{}.
\newblock \showarticletitle{{Migrating Industrial Systems towards Software
  Product Lines: Experiences and Observations through Case Studies}}.
\newblock \bibinfo{journal}{\emph{2008 34th Euromicro Conference Software
  Engineering and Advanced Applications}} (\bibinfo{year}{2008}).
\newblock


\bibitem[Cafeo et~al\mbox{.}(2016)]%
        {CafeoA2016}
\bibfield{author}{\bibinfo{person}{Bruno~B.P. Cafeo}, \bibinfo{person}{Elder
  Cirilo}, \bibinfo{person}{Alessandro Garcia}, \bibinfo{person}{Francisco
  Dantas}, {and} \bibinfo{person}{Jaejoon Lee}.}
  \bibinfo{year}{2016}\natexlab{}.
\newblock \showarticletitle{{Feature dependencies as change propagators: An
  exploratory study of software product lines}}.
\newblock \bibinfo{journal}{\emph{Information and Software Technology}}
  \bibinfo{volume}{69} (\bibinfo{year}{2016}), \bibinfo{pages}{37--49}.
\newblock


\bibitem[Clements and Northrop(2001)]%
        {Clements2001}
\bibfield{author}{\bibinfo{person}{Paul Clements} {and} \bibinfo{person}{Linda
  Northrop}.} \bibinfo{year}{2001}\natexlab{}.
\newblock \bibinfo{booktitle}{\emph{Software Product Lines: Practices and
  Patterns}}.
\newblock \bibinfo{publisher}{Addison-Wesley}, \bibinfo{address}{Boston, MA,
  USA}.
\newblock


\bibitem[Cordy(2006)]%
        {Cordy2006}
\bibfield{author}{\bibinfo{person}{James~R. Cordy}.}
  \bibinfo{year}{2006}\natexlab{}.
\newblock \showarticletitle{The TXL Source Transformation Language}.
\newblock \bibinfo{journal}{\emph{Sci. Comput. Program.}} \bibinfo{volume}{61},
  \bibinfo{number}{3} (\bibinfo{year}{2006}), \bibinfo{pages}{190--210}.
\newblock


\bibitem[{Cornelissen} et~al\mbox{.}(2009)]%
        {Cornelissen2009}
\bibfield{author}{\bibinfo{person}{B. {Cornelissen}}, \bibinfo{person}{A.
  {Zaidman}}, \bibinfo{person}{A. {van Deursen}}, \bibinfo{person}{L.
  {Moonen}}, {and} \bibinfo{person}{R. {Koschke}}.}
  \bibinfo{year}{2009}\natexlab{}.
\newblock \showarticletitle{A Systematic Survey of Program Comprehension
  through Dynamic Analysis}.
\newblock \bibinfo{journal}{\emph{IEEE Transactions on Software Engineering}}
  \bibinfo{volume}{35}, \bibinfo{number}{5} (\bibinfo{year}{2009}),
  \bibinfo{pages}{684--702}.
\newblock


\bibitem[Dubinsky et~al\mbox{.}(2013)]%
        {Dubinsky2013}
\bibfield{author}{\bibinfo{person}{Yael Dubinsky}, \bibinfo{person}{Julia
  Rubin}, \bibinfo{person}{Thorsten Berger}, \bibinfo{person}{Slawomir
  Duszynski}, \bibinfo{person}{Martin Becker}, {and} \bibinfo{person}{Krzysztof
  Czarnecki}.} \bibinfo{year}{2013}\natexlab{}.
\newblock \showarticletitle{An Exploratory Study of Cloning in Industrial
  Software Product Lines}. In \bibinfo{booktitle}{\emph{Proceedings of the 2013
  17th European Conference on Software Maintenance and Reengineering}}
  \emph{(\bibinfo{series}{CSMR ’13})}. \bibinfo{publisher}{IEEE Computer
  Society}, \bibinfo{address}{USA}, \bibinfo{pages}{25–34}.
\newblock
\showISBNx{9780769549484}


\bibitem[{Fischer} et~al\mbox{.}(2015)]%
        {Fischer2015}
\bibfield{author}{\bibinfo{person}{S. {Fischer}}, \bibinfo{person}{L.
  {Linsbauer}}, \bibinfo{person}{R.~E. {Lopez-Herrejon}}, {and}
  \bibinfo{person}{A. {Egyed}}.} \bibinfo{year}{2015}\natexlab{}.
\newblock \showarticletitle{The ECCO Tool: Extraction and Composition for
  Clone-and-Own}. In \bibinfo{booktitle}{\emph{2015 IEEE/ACM 37th IEEE
  International Conference on Software Engineering}}, Vol.~\bibinfo{volume}{2}.
  \bibinfo{pages}{665--668}.
\newblock


\bibitem[Gacek and Anastasopoules(2001)]%
        {Gacek2001}
\bibfield{author}{\bibinfo{person}{Critina Gacek} {and}
  \bibinfo{person}{Michalis Anastasopoules}.} \bibinfo{year}{2001}\natexlab{}.
\newblock \showarticletitle{Implementing Product Line Variabilities}
  \emph{(\bibinfo{series}{SSR '01})}. \bibinfo{publisher}{Association for
  Computing Machinery}, \bibinfo{address}{New York, NY, USA},
  \bibinfo{pages}{109–117}.
\newblock
\showISBNx{1581133588}


\bibitem[Gelman(2005)]%
        {Gelman2005}
\bibfield{author}{\bibinfo{person}{Andrew Gelman}.}
  \bibinfo{year}{2005}\natexlab{}.
\newblock \showarticletitle{Analysis of variance: Why it is more important than
  ever?}
\newblock \bibinfo{journal}{\emph{Quality Engineering}}  \bibinfo{volume}{51}
  (\bibinfo{year}{2005}), \bibinfo{pages}{295--300}.
\newblock


\bibitem[Harman et~al\mbox{.}(2013)]%
        {Harman2013}
\bibfield{author}{\bibinfo{person}{Mark Harman}, \bibinfo{person}{William~B
  Langdon}, {and} \bibinfo{person}{Westley Weimer}.}
  \bibinfo{year}{2013}\natexlab{}.
\newblock \showarticletitle{{Genetic Programming for Reverse Engineering}}.
\newblock  (\bibinfo{year}{2013}), \bibinfo{pages}{1--10}.
\newblock
\showISBNx{9781479929313}


\bibitem[Hlad et~al\mbox{.}(2021)]%
        {hlad2021}
\bibfield{author}{\bibinfo{person}{Nicolas Hlad}, \bibinfo{person}{Seriai
  Abdelhak-Djamel}, {and} \bibinfo{person}{Dony Christophe}.}
  \bibinfo{year}{2021}\natexlab{}.
\newblock \bibinfo{title}{IsiSPL: Toward an Automated Reactive Approach to
  build Software Product Lines}.
\newblock
\newblock
\showeprint[arxiv]{2107.00552}~[cs.SE]


\bibitem[Ilyas and Elkhalifa(2016)]%
        {Ilyas2016}
\bibfield{author}{\bibinfo{person}{Bilal Ilyas} {and} \bibinfo{person}{Islam
  Elkhalifa}.} \bibinfo{year}{2016}\natexlab{}.
\newblock \showarticletitle{Static Code Analysis: A Systematic Literature
  Review and an Industrial Survey}.
\newblock


\bibitem[J\'{e}z\'{e}quel et~al\mbox{.}(2022)]%
        {Jezequel2022}
\bibfield{author}{\bibinfo{person}{Jean-Marc J\'{e}z\'{e}quel},
  \bibinfo{person}{J\"{o}rg Kienzle}, {and} \bibinfo{person}{Mathieu Acher}.}
  \bibinfo{year}{2022}\natexlab{}.
\newblock \showarticletitle{From Feature Models to Feature Toggles in
  Practice}. In \bibinfo{booktitle}{\emph{Proceedings of the 26th ACM
  International Systems and Software Product Line Conference - Volume A}}
  (Graz, Austria) \emph{(\bibinfo{series}{SPLC '22})}.
  \bibinfo{publisher}{Association for Computing Machinery},
  \bibinfo{address}{New York, NY, USA}, \bibinfo{pages}{234–244}.
\newblock


\bibitem[Kang et~al\mbox{.}(1990)]%
        {Kang1990}
\bibfield{author}{\bibinfo{person}{K.~C. Kang}, \bibinfo{person}{S.~G. Cohen},
  \bibinfo{person}{J.~A. Hess}, \bibinfo{person}{W.~E. Novak}, {and}
  \bibinfo{person}{A.~S. Peterson}.} \bibinfo{year}{1990}\natexlab{}.
\newblock \bibinfo{booktitle}{\emph{Feature-Oriented Domain Analysis (FODA)
  Feasibility Study}}.
\newblock \bibinfo{type}{{T}echnical {R}eport}.
  \bibinfo{institution}{Carnegie-Mellon University Software Engineering
  Institute}.
\newblock


\bibitem[K\"{a}stner et~al\mbox{.}(2008)]%
        {Kastner2008B}
\bibfield{author}{\bibinfo{person}{Christian K\"{a}stner},
  \bibinfo{person}{Sven Apel}, {and} \bibinfo{person}{Martin Kuhlemann}.}
  \bibinfo{year}{2008}\natexlab{}.
\newblock \showarticletitle{Granularity in Software Product Lines}
  \emph{(\bibinfo{series}{ICSE '08})}. \bibinfo{publisher}{Association for
  Computing Machinery}, \bibinfo{address}{New York, NY, USA},
  \bibinfo{pages}{311–320}.
\newblock


\bibitem[Kernighan and Pike(1983)]%
        {Kernighan1983}
\bibfield{author}{\bibinfo{person}{Brian~W. Kernighan} {and}
  \bibinfo{person}{Rob Pike}.} \bibinfo{year}{1983}\natexlab{}.
\newblock \bibinfo{booktitle}{\emph{The UNIX Programming Environment}}.
\newblock \bibinfo{publisher}{Prentice Hall Professional Technical Reference}.
\newblock


\bibitem[Krueger(2002a)]%
        {Krueger2001}
\bibfield{author}{\bibinfo{person}{Charles~W. Krueger}.}
  \bibinfo{year}{2002}\natexlab{a}.
\newblock \showarticletitle{Easing the Transition to Software Mass
  Customization}. In \bibinfo{booktitle}{\emph{Revised Papers from the 4th
  International Workshop on Software Product-Family Engineering}}
  \emph{(\bibinfo{series}{PFE '01})}. \bibinfo{publisher}{Springer-Verlag},
  \bibinfo{address}{London, UK, UK}, \bibinfo{pages}{282--293}.
\newblock


\bibitem[Krueger(2002b)]%
        {Krueger2002}
\bibfield{author}{\bibinfo{person}{Charles~W. Krueger}.}
  \bibinfo{year}{2002}\natexlab{b}.
\newblock \showarticletitle{Easing the Transition to Software Mass
  Customization}. In \bibinfo{booktitle}{\emph{Revised Papers from the 4th
  International Workshop on Software Product-Family Engineering}}
  \emph{(\bibinfo{series}{PFE '01})}. \bibinfo{publisher}{Springer-Verlag},
  \bibinfo{address}{London, UK, UK}, \bibinfo{pages}{282--293}.
\newblock


\bibitem[Kr{\"{u}}ger et~al\mbox{.}(2018)]%
        {Kruger2018}
\bibfield{author}{\bibinfo{person}{Jacob Kr{\"{u}}ger},
  \bibinfo{person}{Thorsten Berger}, \bibinfo{person}{Thomas Leich}, {and}
  \bibinfo{person}{" Features}.} \bibinfo{year}{2018}\natexlab{}.
\newblock \showarticletitle{{Features and How to Find Them: A Survey of Manual
  Feature Location Feature location; systematic literature review; reverse
  variability engineering; feature identification; feature mapping}}.
\newblock \bibinfo{journal}{\emph{Software Engineering for Variability
  Intensive Systems: Foundations and Applications}} (\bibinfo{year}{2018}).
\newblock


\bibitem[Kr\"{u}ger et~al\mbox{.}(2020)]%
        {Kruger2020}
\bibfield{author}{\bibinfo{person}{Jacob Kr\"{u}ger},
  \bibinfo{person}{Sebastian Krieter}, \bibinfo{person}{Gunter Saake}, {and}
  \bibinfo{person}{Thomas Leich}.} \bibinfo{year}{2020}\natexlab{}.
\newblock \showarticletitle{EXtracting Product Lines from VAriaNTs (EXPLANT)}
  \emph{(\bibinfo{series}{VAMOS '20})}. \bibinfo{publisher}{Association for
  Computing Machinery}, \bibinfo{address}{New York, NY, USA}, Article
  \bibinfo{articleno}{13}, \bibinfo{numpages}{2}~pages.
\newblock


\bibitem[{Kästner} et~al\mbox{.}(2014)]%
        {Kastner2014}
\bibfield{author}{\bibinfo{person}{C. {Kästner}}, \bibinfo{person}{A.
  {Dreiling}}, {and} \bibinfo{person}{K. {Ostermann}}.}
  \bibinfo{year}{2014}\natexlab{}.
\newblock \showarticletitle{Variability Mining: Consistent Semi-automatic
  Detection of Product-Line Features}.
\newblock \bibinfo{journal}{\emph{IEEE Transactions on Software Engineering}}
  \bibinfo{volume}{40}, \bibinfo{number}{1} (\bibinfo{year}{2014}),
  \bibinfo{pages}{67--82}.
\newblock


\bibitem[Kästner et~al\mbox{.}(2008)]%
        {Kastner2008}
\bibfield{author}{\bibinfo{person}{Christian Kästner},
  \bibinfo{person}{Salvador Trujillo}, {and} \bibinfo{person}{Sven Apel}.}
  \bibinfo{year}{2008}\natexlab{}.
\newblock \showarticletitle{Visualizing Software Product Line Variabilities in
  Source Code}. In \bibinfo{booktitle}{\emph{In Proc. SPLC Workshop on
  Visualization in Software Product Line Engineering (ViSPLE}}.
\newblock


\bibitem[Laguna and Crespo(2013)]%
        {Laguna2013}
\bibfield{author}{\bibinfo{person}{Miguel~A. Laguna} {and}
  \bibinfo{person}{Yania Crespo}.} \bibinfo{year}{2013}\natexlab{}.
\newblock \showarticletitle{A systematic mapping study on software product line
  evolution: From legacy system reengineering to product line refactoring}.
\newblock \bibinfo{journal}{\emph{Science of Computer Programming}}
  \bibinfo{volume}{78}, \bibinfo{number}{8} (\bibinfo{year}{2013}),
  \bibinfo{pages}{1010 -- 1034}.
\newblock


\bibitem[Liebig et~al\mbox{.}(2010)]%
        {Liebig2010}
\bibfield{author}{\bibinfo{person}{Jorg Liebig}, \bibinfo{person}{Sven Apel},
  \bibinfo{person}{Christian Lengauer}, \bibinfo{person}{Christian Kästner},
  {and} \bibinfo{person}{Michael Schulze}.} \bibinfo{year}{2010}\natexlab{}.
\newblock \showarticletitle{An analysis of the variability in forty
  preprocessor-based software product lines}. In \bibinfo{booktitle}{\emph{2010
  ACM/IEEE 32nd International Conference on Software Engineering}},
  Vol.~\bibinfo{volume}{1}. \bibinfo{pages}{105--114}.
\newblock


\bibitem[Linden et~al\mbox{.}(2007)]%
        {Linden2007}
\bibfield{author}{\bibinfo{person}{Frank J. van~der Linden},
  \bibinfo{person}{Klaus Schmid}, {and} \bibinfo{person}{Eelco Rommes}.}
  \bibinfo{year}{2007}\natexlab{}.
\newblock \bibinfo{booktitle}{\emph{Software Product Lines in Action: The Best
  Industrial Practice in Product Line Engineering}}.
\newblock \bibinfo{publisher}{Springer-Verlag New York, Inc.},
  \bibinfo{address}{Secaucus, NJ, USA}.
\newblock


\bibitem[Liu and Mao(2018)]%
        {Liu2018}
\bibfield{author}{\bibinfo{person}{Lupeng Liu} {and} \bibinfo{person}{Xiaoguang
  Mao}.} \bibinfo{year}{2018}\natexlab{}.
\newblock \showarticletitle{{A Study on Code Transplantation Technique based on
  Program Slicing}}.
\newblock  \bibinfo{volume}{161}, \bibinfo{number}{Tlicsc}
  (\bibinfo{year}{2018}), \bibinfo{pages}{294--298}.
\newblock


\bibitem[Lotufo et~al\mbox{.}(2010)]%
        {Lotufo2010}
\bibfield{author}{\bibinfo{person}{Rafael Lotufo}, \bibinfo{person}{Steven
  She}, \bibinfo{person}{Thorsten Berger}, \bibinfo{person}{Krzysztof
  Czarnecki}, {and} \bibinfo{person}{Andrzej Wasowski}.}
  \bibinfo{year}{2010}\natexlab{}.
\newblock \showarticletitle{Evolution of the Linux Kernel Variability Model}
  \emph{(\bibinfo{series}{SPLC'10})}. \bibinfo{publisher}{Springer-Verlag},
  \bibinfo{address}{Berlin, Heidelberg}, \bibinfo{pages}{136–150}.
\newblock
\showISBNx{3642155782}


\bibitem[Mahmood et~al\mbox{.}(2021)]%
        {Mahmood2021}
\bibfield{author}{\bibinfo{person}{Wardah Mahmood}, \bibinfo{person}{Daniel
  Str{\"u}ber}, \bibinfo{person}{Thorsten Berger}, \bibinfo{person}{Ralf
  L{\"a}mmel}, {and} \bibinfo{person}{Mukelabai Mukelabai}.}
  \bibinfo{year}{2021}\natexlab{}.
\newblock \showarticletitle{Seamless Variability Management with the Virtual
  Platform}.
\newblock \bibinfo{journal}{\emph{2021 IEEE/ACM 43rd International Conference
  on Software Engineering (ICSE)}} (\bibinfo{year}{2021}),
  \bibinfo{pages}{1658--1670}.
\newblock


\bibitem[Martinez et~al\mbox{.}(2015)]%
        {Martinez2015}
\bibfield{author}{\bibinfo{person}{Jabier Martinez}, \bibinfo{person}{Tewfik
  Ziadi}, \bibinfo{person}{Tegawend\'{e}~F. Bissyand\'{e}},
  \bibinfo{person}{Jacques Klein}, {and} \bibinfo{person}{Yves Le~Traon}.}
  \bibinfo{year}{2015}\natexlab{}.
\newblock \showarticletitle{Bottom-up Adoption of Software Product Lines: A
  Generic and Extensible Approach}. In \bibinfo{booktitle}{\emph{Proceedings of
  the 19th International Conference on Software Product Line}} (Nashville,
  Tennessee) \emph{(\bibinfo{series}{SPLC ’15})}.
  \bibinfo{publisher}{Association for Computing Machinery},
  \bibinfo{address}{New York, NY, USA}, \bibinfo{pages}{101–110}.
\newblock


\bibitem[Meinicke et~al\mbox{.}(2020)]%
        {Meinicke2020}
\bibfield{author}{\bibinfo{person}{Jens Meinicke}, \bibinfo{person}{Chu-Pan
  Wong}, \bibinfo{person}{Bogdan Vasilescu}, {and} \bibinfo{person}{Christian
  K\"{a}stner}.} \bibinfo{year}{2020}\natexlab{}.
\newblock \showarticletitle{Exploring Differences and Commonalities between
  Feature Flags and Configuration Options} \emph{(\bibinfo{series}{ICSE-SEIP
  '20})}. \bibinfo{publisher}{Association for Computing Machinery},
  \bibinfo{address}{New York, NY, USA}, \bibinfo{pages}{233–242}.
\newblock


\bibitem[Northrop and C.(2012)]%
        {Northrop2012}
\bibfield{author}{\bibinfo{person}{Linda~M. Northrop} {and}
  \bibinfo{person}{Clements~Paul C.}} \bibinfo{year}{2012}\natexlab{}.
\newblock \showarticletitle{{A Framework for Software Product Line Practice
  version 5.0. technical report}}.
\newblock \bibinfo{journal}{\emph{Software Engineering Institute}}
  (\bibinfo{year}{2012}), \bibinfo{pages}{258}.
\newblock
\urldef\tempurl%
\url{http://www.sei.cmu.edu/productlines/frame{\_}report/index.html}
\showURL{%
\tempurl}


\bibitem[Petke et~al\mbox{.}(2018)]%
        {Petke18}
\bibfield{author}{\bibinfo{person}{Justyna Petke},
  \bibinfo{person}{Saemundur~O. Haraldsson}, \bibinfo{person}{Mark Harman},
  \bibinfo{person}{William~B. Langdon}, \bibinfo{person}{David~Robert White},
  {and} \bibinfo{person}{John~R. Woodward}.} \bibinfo{year}{2018}\natexlab{}.
\newblock \showarticletitle{Genetic Improvement of Software: {A} Comprehensive
  Survey}.
\newblock \bibinfo{journal}{\emph{{IEEE} Trans. Evol. Comput.}}
  \bibinfo{volume}{22}, \bibinfo{number}{3} (\bibinfo{year}{2018}),
  \bibinfo{pages}{415--432}.
\newblock


\bibitem[Petke et~al\mbox{.}(2014)]%
        {Petke2014}
\bibfield{author}{\bibinfo{person}{Justyna Petke}, \bibinfo{person}{Mark
  Harman}, \bibinfo{person}{William~B. Langdon}, {and} \bibinfo{person}{Westley
  Weimer}.} \bibinfo{year}{2014}\natexlab{}.
\newblock \showarticletitle{Using Genetic Improvement and Code Transplants to
  Specialise a C++ Program to a Problem Class}. In
  \bibinfo{booktitle}{\emph{Genetic Programming}},
  \bibfield{editor}{\bibinfo{person}{Miguel Nicolau},
  \bibinfo{person}{Krzysztof Krawiec}, \bibinfo{person}{Malcolm~I. Heywood},
  \bibinfo{person}{Mauro Castelli}, \bibinfo{person}{Pablo
  Garc{\'i}a-S{\'a}nchez}, \bibinfo{person}{Juan~J. Merelo},
  \bibinfo{person}{Victor~M. Rivas~Santos}, {and} \bibinfo{person}{Kevin Sim}}
  (Eds.). \bibinfo{publisher}{Springer Berlin Heidelberg},
  \bibinfo{address}{Berlin, Heidelberg}, \bibinfo{pages}{137--149}.
\newblock


\bibitem[{Petke} et~al\mbox{.}(2018)]%
        {Petke2018}
\bibfield{author}{\bibinfo{person}{J. {Petke}}, \bibinfo{person}{M. {Harman}},
  \bibinfo{person}{W.~B. {Langdon}}, {and} \bibinfo{person}{W. {Weimer}}.}
  \bibinfo{year}{2018}\natexlab{}.
\newblock \showarticletitle{Specialising Software for Different Downstream
  Applications Using Genetic Improvement and Code Transplantation}.
\newblock \bibinfo{journal}{\emph{IEEE Transactions on Software Engineering}}
  \bibinfo{volume}{44}, \bibinfo{number}{6} (\bibinfo{date}{June}
  \bibinfo{year}{2018}), \bibinfo{pages}{574--594}.
\newblock


\bibitem[Pohl et~al\mbox{.}(2005)]%
        {Pohl2005}
\bibfield{author}{\bibinfo{person}{Klaus Pohl}, \bibinfo{person}{G\"{u}nter
  B\"{o}ckle}, {and} \bibinfo{person}{Frank J. van~der Linden}.}
  \bibinfo{year}{2005}\natexlab{}.
\newblock \bibinfo{booktitle}{\emph{Software Product Line Engineering:
  Foundations, Principles and Techniques}}.
\newblock \bibinfo{publisher}{Springer-Verlag New York, Inc.},
  \bibinfo{address}{Secaucus, NJ, USA}.
\newblock


\bibitem[Rahman et~al\mbox{.}(2016)]%
        {Rahman2016}
\bibfield{author}{\bibinfo{person}{Md~Tajmilur Rahman},
  \bibinfo{person}{Louis-Philippe Querel}, \bibinfo{person}{Peter~C. Rigby},
  {and} \bibinfo{person}{Bram Adams}.} \bibinfo{year}{2016}\natexlab{}.
\newblock \showarticletitle{Feature Toggles: Practitioner Practices and a Case
  Study}. In \bibinfo{booktitle}{\emph{Proceedings of the 13th International
  Conference on Mining Software Repositories}} (Austin, Texas)
  \emph{(\bibinfo{series}{MSR '16})}. \bibinfo{publisher}{Association for
  Computing Machinery}, \bibinfo{address}{New York, NY, USA},
  \bibinfo{pages}{201–211}.
\newblock


\bibitem[Ribeiro et~al\mbox{.}(2011)]%
        {Ribeiro2011}
\bibfield{author}{\bibinfo{person}{M\'{a}rcio Ribeiro}, \bibinfo{person}{Felipe
  Queiroz}, \bibinfo{person}{Paulo Borba}, \bibinfo{person}{T\'{a}rsis
  Tol\^{e}do}, \bibinfo{person}{Claus Brabrand}, {and}
  \bibinfo{person}{S{\'e}rgio Soares}.} \bibinfo{year}{2011}\natexlab{}.
\newblock \showarticletitle{On the Impact of Feature Dependencies when
  Maintaining Preprocessor-based Software Product Lines}.
\newblock \bibinfo{journal}{\emph{SIGPLAN Not.}} \bibinfo{volume}{47},
  \bibinfo{number}{3} (\bibinfo{date}{Oct.} \bibinfo{year}{2011}),
  \bibinfo{pages}{23--32}.
\newblock
\showISSN{0362-1340}


\bibitem[Roy(2009)]%
        {Roy2009}
\bibfield{author}{\bibinfo{person}{Chanchal~K. Roy}.}
  \bibinfo{year}{2009}\natexlab{}.
\newblock \emph{\bibinfo{title}{Detection and Analysis of Near-miss Software
  Clones}}.
\newblock \bibinfo{thesistype}{Ph.\,D. Dissertation}.
  \bibinfo{address}{Kingston, Ont., Canada, Canada}.
\newblock
\showISBNx{978-0-494-65337-1}
\newblock
\shownote{AAINR65337}.


\bibitem[SHAPIRO and WILK(1965)]%
        {SHAPIRO1965}
\bibfield{author}{\bibinfo{person}{S.~S. SHAPIRO} {and} \bibinfo{person}{M.~B.
  WILK}.} \bibinfo{year}{1965}\natexlab{}.
\newblock \showarticletitle{{An analysis of variance test for normality
  (complete samples)†}}.
\newblock \bibinfo{journal}{\emph{Biometrika}} \bibinfo{volume}{52},
  \bibinfo{number}{3-4} (\bibinfo{date}{12} \bibinfo{year}{1965}),
  \bibinfo{pages}{591--611}.
\newblock


\bibitem[Sidiroglou-Douskos et~al\mbox{.}(2017)]%
        {Sidiroglou2017}
\bibfield{author}{\bibinfo{person}{Stelios Sidiroglou-Douskos},
  \bibinfo{person}{Eric Lahtinen}, \bibinfo{person}{Anthony Eden},
  \bibinfo{person}{Fan Long}, {and} \bibinfo{person}{Martin Rinard}.}
  \bibinfo{year}{2017}\natexlab{}.
\newblock \showarticletitle{CodeCarbonCopy}. In
  \bibinfo{booktitle}{\emph{Proceedings of the 2017 11th Joint Meeting on
  Foundations of Software Engineering}} (Paderborn, Germany)
  \emph{(\bibinfo{series}{ESEC/FSE 2017})}. \bibinfo{address}{New York, NY,
  USA}, \bibinfo{pages}{95–105}.
\newblock


\bibitem[Souza et~al\mbox{.}(2023)]%
        {ProjectWebpage}
\bibfield{author}{\bibinfo{person}{Leandro Souza}, \bibinfo{person}{Earl Barr},
  \bibinfo{person}{Justyna Petke}, \bibinfo{person}{Eduardo Almeida}, {and}
  \bibinfo{person}{Paulo Neto}.} \bibinfo{year}{2023}\natexlab{}.
\newblock \bibinfo{title}{The project: software product line engineering via
  automated software transplantation}.
\newblock
  \bibinfo{howpublished}{\url{https://autotransplantation-spl.github.io/foundry.github.io/}}.
\newblock
\newblock
\shownote{Accessed: 2023-04-25}.


\bibitem[Tartler et~al\mbox{.}(2011)]%
        {Tartler2011}
\bibfield{author}{\bibinfo{person}{Reinhard Tartler}, \bibinfo{person}{Daniel
  Lohmann}, \bibinfo{person}{Julio Sincero}, {and} \bibinfo{person}{Wolfgang
  Schr\"{o}der-Preikschat}.} \bibinfo{year}{2011}\natexlab{}.
\newblock \showarticletitle{Feature Consistency in Compile-time-configurable
  System Software: Facing the Linux 10,000 Feature Problem}. In
  \bibinfo{booktitle}{\emph{Proceedings of the Sixth Conference on Computer
  Systems}} (Salzburg, Austria) \emph{(\bibinfo{series}{EuroSys '11})}.
  \bibinfo{publisher}{ACM}, \bibinfo{address}{New York, NY, USA},
  \bibinfo{pages}{47--60}.
\newblock


\bibitem[van Heesch(2018)]%
        {Doxygen2018}
\bibfield{author}{\bibinfo{person}{D. van Heesch}.}
  \bibinfo{year}{2018}\natexlab{}.
\newblock \bibinfo{title}{Doxygen: Source code documentation generator tool}.
\newblock
\newblock
\urldef\tempurl%
\url{http://www.stack.nl/dimitri/doxygen/}
\showURL{%
\tempurl}


\bibitem[Wang et~al\mbox{.}(2018)]%
        {Wang2018}
\bibfield{author}{\bibinfo{person}{Shangwen Wang}, \bibinfo{person}{Xiaoguang
  Mao}, {and} \bibinfo{person}{Yue Yu}.} \bibinfo{year}{2018}\natexlab{}.
\newblock \showarticletitle{An Initial Step Towards Organ Transplantation Based
  on GitHub Repository}.
\newblock \bibinfo{journal}{\emph{IEEE Access}}  \bibinfo{volume}{6}
  (\bibinfo{year}{2018}), \bibinfo{pages}{59268--59281}.
\newblock


\bibitem[Wohlin et~al\mbox{.}(2012)]%
        {Wohlin2012b}
\bibfield{author}{\bibinfo{person}{Claes Wohlin}, \bibinfo{person}{Per
  Runeson}, \bibinfo{person}{Martin Hst}, \bibinfo{person}{Magnus~C. Ohlsson},
  \bibinfo{person}{Bjrn Regnell}, {and} \bibinfo{person}{Anders Wessln}.}
  \bibinfo{year}{2012}\natexlab{}.
\newblock \bibinfo{booktitle}{\emph{Experimentation in Software Engineering}}.
\newblock \bibinfo{publisher}{Springer Publishing Company, Incorporated}.
\newblock


\bibitem[Xu et~al\mbox{.}(2015)]%
        {Xu2015}
\bibfield{author}{\bibinfo{person}{Tianyin Xu}, \bibinfo{person}{Long Jin},
  \bibinfo{person}{Xuepeng Fan}, \bibinfo{person}{Yuanyuan Zhou},
  \bibinfo{person}{Shankar Pasupathy}, {and} \bibinfo{person}{Rukma
  Talwadker}.} \bibinfo{year}{2015}\natexlab{}.
\newblock \showarticletitle{Hey, You Have given Me Too Many Knobs!:
  Understanding and Dealing with over-Designed Configuration in System
  Software}. In \bibinfo{booktitle}{\emph{Proceedings of the 2015 10th Joint
  Meeting on Foundations of Software Engineering}} (Bergamo, Italy)
  \emph{(\bibinfo{series}{ESEC/FSE 2015})}. \bibinfo{publisher}{Association for
  Computing Machinery}, \bibinfo{address}{New York, NY, USA},
  \bibinfo{pages}{307–319}.
\newblock


\bibitem[Yoshimura et~al\mbox{.}(2006)]%
        {Yoshimura2006}
\bibfield{author}{\bibinfo{person}{Kentaro Yoshimura},
  \bibinfo{person}{Dharmalingam Ganesan}, {and} \bibinfo{person}{Dirk Muthig}.}
  \bibinfo{year}{2006}\natexlab{}.
\newblock \showarticletitle{Defining a Strategy to Introduce a Software Product
  Line Using Existing Embedded Systems}. In
  \bibinfo{booktitle}{\emph{Proceedings of the 6th ACM \& IEEE International
  Conference on Embedded Software}} (Seoul, Korea)
  \emph{(\bibinfo{series}{EMSOFT ’06})}. \bibinfo{publisher}{Association for
  Computing Machinery}, \bibinfo{address}{New York, NY, USA},
  \bibinfo{pages}{63–72}.
\newblock


\end{thebibliography}
